\documentclass[12pt]{svjour2}

\smartqed

\usepackage{amssymb}
\usepackage{graphicx}

\journalname{J Stat Phys}

\begin{document}

\title{Attraction of like-charged walls with counterions only: Exact 
results for the 2D cylinder geometry}

\titlerunning{Attraction of like-charged walls with counterions only}

\author{Ladislav \v{S}amaj}

\institute{Institute of Physics, Slovak Academy of Sciences, 
D\'ubravsk\'a cesta 9, SK-84511 Bratislava, Slovakia \\
\email{Ladislav.Samaj@savba.sk}}

\date{Received:  / Accepted: }

\maketitle

\begin{abstract}
We study a 2D system of identical mobile particles on the surface of 
a cylinder of finite length $d$ and circumference $W$, immersed in
a medium of dielectric constant $\varepsilon$. 
The two end-circles of the cylinder are like-charged with the fixed uniform 
charge densities, the particles of opposite charge $-e$ ($e$ being 
the elementary charge) are coined as ``counterions''; 
the system as a whole is electroneutral.
Such a geometry is well defined also for finite numbers of counterions $N$.
Our task is to derive an effective interaction between the end-circles
mediated by the counterions in thermal equilibrium at the inverse temperature 
$\beta$.
The exact solution of the system at the free-fermion coupling 
$\Gamma \equiv \beta e^2/\varepsilon =2$ is used to test the convergence
of the pressure as the (even) number of particles increases from 
$N=2$ to $\infty$.
The pressure as a function of distance $d$ is always positive 
(effective repulsion between the like-charged circles), decaying
monotonously; the numerical results for $N=8$ counterions are 
very close to those in the thermodynamic limit $N\to\infty$.
For the couplings $\Gamma=2\gamma$ with $\gamma=1,2,\ldots$, there exists
a mapping of the continuous two-dimensional (2D) Coulomb system with $N$ 
particles onto the one-dimensional (1D) lattice model of $N$ sites with 
interacting sets of anticommuting variables.
This allows one to treat exactly the density profile, two-body density and
the pressure for the couplings $\Gamma=4$ and $6$, up to $N=8$ particles.
Our main finding is that the pressure becomes negative at large enough 
distances $d$ if and only if both like-charged walls carry a nonzero charge 
density.
This indicates a like-attraction in the thermodynamic limit $N\to\infty$ 
as well, starting from a relatively weak coupling constant 
$\Gamma$ in between 2 and 4. 
As a by-product of the formalism, we derive specific sum rules which have 
direct impact on characteristics of the long-range decay of 2D two-body 
densities along the two walls. 


\end{abstract}

\renewcommand{\theequation}{1.\arabic{equation}}
\setcounter{equation}{0}

\section{Introduction} \label{Sec:1}
The study of equilibrium statistical mechanics of classical (i.e. nonquantum) 
systems of particles interacting pairwisely by Coulomb potential is 
of particular importance in condensed matter and soft matter physics.
In the real $3$-dimensional (3D) space of practical interest, the Coulomb 
potential in vacuum of dielectric constant $\varepsilon=1$ has in Gauss
units the standard form $\phi({\bf r}) = 1/r$ with $r$ being 
the modulus of ${\bf r}$. 
The definition of the Coulomb potential can be extended to any dimension 
$\nu=1,2,\ldots$ as the solution of the Poisson equation
\begin{equation} \label{PE}
\Delta \phi({\bf r}) = - s_{\nu} \delta({\bf r}) ,
\end{equation}
where $s_{\nu}=2\pi^{\nu/2}/\Gamma(\nu/2)$ ($\Gamma$ being the Gamma 
function) is the surface area of the $\nu$-dimensional unit sphere.
In an infinite space, the solution of (\ref{PE}), subject to the boundary
condition $\nabla \phi({\bf r}) \to 0$ as $r\to\infty$, reads as
$r$ in 1D, $-\ln(r/L)$ ($L$ is a free length scale) in 2D and 
$r^{2-\nu}/(\nu-2)$ in spatial dimensions $\nu\ge 3$.
The Fourier component of the Coulomb potential exhibits the singular $1/k^2$ 
behavior in any dimension; this maintains many generic properties of 
3D Coulomb systems like screening \cite{Martin88}. 

In one-component Coulomb models, the system of mobile (pointlike) particles of 
the same (say elementary) charge $-e$ is neutralized by a fixed ``background'' 
charge density.
The most known system of this kind is the jellium model of real materials
in which the homogeneous background charge density of heavy nucleus ions
is spread over the whole space of the domain mobile electrons are confined 
to \cite{Baus80}.
Since due to the electroneutrality requirement the particle number is 
proportional to the domain's volume, the system is ``dense'' and therefore 
exhibits good screening properties in thermal equilibrium, i.e., the particle
correlation functions exhibit a short-range, usually exponential, decay 
at asymptotically large distances.
The 1D jellium model, treated by using a transfer matrix method
\cite{Lenard61} and a generating function method \cite{Edwards62}, 
is exactly solvable for any temperature and particle density. 
The system exhibits the translational symmetry breaking of the particle 
number density which oscillates periodically in the thermodynamic limit 
\cite{Baxter63,Kunz74}. 
Boundary effects are important in 1D \cite{Dean98}.
In 2D characterized by the logarithmic Coulomb potential, the relevant 
coupling constant is $\Gamma\equiv\beta e^2/\varepsilon$ where $\beta$
is the inverse temperature and $\varepsilon$ is the dielectric constant of
the medium the particles are immersed in.
These systems are especially important because they are exactly solvable, 
besides the mean-field $\Gamma\to 0$ limit, also at a special finite 
temperature.
The exact solution of the 2D jellium model at the ``free-fermion'' coupling 
$\Gamma=2$ involves the bulk case \cite{Alastuey81,Jancovici81} as well 
as semi-infinite and fully finite geometries, see reviews 
\cite{Forrester98,Jancovici92}.

A series of works was devoted to the study of thermal equilibrium 
of 2D one-component Coulomb systems for a series of couplings 
$\Gamma=2\gamma$ where $\gamma=1,2,3,\ldots$ is a (positive) integer.
There are two basic approaches how to express integer powers of 
the Vandermonde determinants. 
The method using a mapping of the 2D Coulomb system onto a 1D lattice 
anticommuting-field theory was initiated in \cite{Samaj95} and subsequently
used in a series of works 
\cite{Samaj00,Samaj04a,Samaj04b,Samaj15,Samaj16b,Samaj17}
dealing with sum rules, finite-size corrections, asymptotic decay of
two-body correlations along domain's boundaries, etc.
Another method using Jack polynomials was developed in \cite{Tellez99,Tellez12}.
The relation between the two methods was established in \cite{Grimaldo15}.

In this paper, another version of the one-component Coulomb systems is 
studied, with the background charge density spread over the boundary 
of the constraining domain. 
Due to the electroneutrality, the number of mobile charges is proportional
to the domain's boundary and the screening properties of the ``sparse'' 
system are not good.
This kind of models describes biological experiments with macromolecules 
(colloids, polyelectrolytes) which are performed in polar solvents
like water.
Through the dissociation of functional surface groups, the surface of 
macromolecule releases micro-ions into the polar solvent, acquiring in 
this way a fixed surface charge density \cite{Andelman06,Levin02}.
Since the charge of micro-ions is opposite to that of the surface charge 
density, they are coined as ``counterions''.
In thermal equilibrium, the charged surface of the macromolecule and 
the surrounding counterions form a neutral entity known as 
the electric double layer \cite{Attard88,Attard96,Gulbrand84,Messina09}.
The effective interaction between two like-charged walls, mediated by 
counterions, is of primary experimental and theoretical interest 
\cite{Hansen00}.
At small enough temperatures, a counter-intuitive attraction of like-charged 
macromolecules was observed experimentally
\cite{Bloomfield91,Dubois98,Kekicheff93,Khan85,Kjellander88,Rau92}
as well as by computer simulations
\cite{Bratko86,Gronbech97,Gulbrand84,Kjellander84};
for more recent numerical and analytical advances in this field, see
reviews \cite{Boroudjerdi05,Levin02,Naji13}. 

For large macromolecules with the surface charge of order of thousands
elementary charges, the curved surface can be replaced by an infinite
rectilinear wall.
Thermal equilibrium of charged surfaces with counterions only is usually
considered in the canonical ensemble at the inverse temperature 
$\beta=1/(k_{\rm B}T)$. 
Two basic geometries are studied. 
In the case of one wall with counterions constrained to the semi-infinite 
(half) space, the particle density profile is of interest.
The particle density at the wall is related to the wall's surface
charge density via the contact-value theorem 
\cite{Blum81,Carnie81,Choquard80,Henderson78,Henderson79}.
To obtain the effective interaction of two parallel walls at distance $d$,
one calculates the pressure, either from the derivative of the free energy
with respect to $d$ or from contact densities.
Since the background charge is confined to the surfaces of the walls,
it stays inside the system when changing infinitesimally $d$ and so, 
in contrast to one-component jellium systems \cite{Choquard80}, there is 
no ambiguity in the definition of the pressure. 
  
From a theoretical point of view, models of charged wall surfaces with 
counterions only are probably the simplest ones to study the equilibrium 
properties of Coulomb fluids.
The weak-coupling (high-temperature, WC) limit is described by the mean-field
Poisson-Boltzmann (PB) approach \cite{Andelman06} and by its systematic 
improvement within the field-theoretical formulation via the loop expansion 
\cite{Attard88,Netz00,Podgornik90}.
In a single pure solvent, two symmetrically charged walls always repel one 
another in the WC limit; this is no longer true for a mixture of polar
solvents when the medium becomes inhomogeneous due to solvation-related
forces \cite{Ben11,Samin11}.

The strong-coupling (low-temperature, SC) limit of the fluid regime 
is more controversial.
Within the virial SC theory put forward by Moreira and Netz
\cite{Moreira00,Moreira01,Moreira02,Netz01}, 
the leading SC term of the counterion density corresponds to a single 
(noninteracting) particle theory in the electric potential of charged wall(s) 
which has been confirmed by Monte Carlo (MC) simulations
\cite{Dean09,Kanduc07,Kanduc08,Moreira00,Moreira01,Moreira02,Naji05}.
Next correction orders in inverse powers of the coupling constant, obtained 
within a virial fugacity expansion, require a renormalization
of infrared divergencies via the electroneutrality condition.
Comparison with MC simulations shows that the first correction term
has the correct functional form in space, but the wrong prefactor.
A dressed-ion version of the virial SC theory was applied to realistic 
Coulomb fluids in the presence of salt \cite{Kanduc10,Kanduc11,Kanduc12};
such an approach has been tested against simulations therein and against
experiments in \cite{Kanduc17}. 

Another type of SC approaches is based on the formation of the classical
Wigner crystal of counterions on the wall surfaces at zero temperature 
\cite{Grosberg02,Levin99,Shklovskii99}.
Based on a harmonic expansion of the interaction energy in particle deviations 
from their ground-state Wigner positions \cite{Samaj11a,Samaj11b}, the
leading single-particle picture of the virial SC approach was reproduced.
The first correction term to the counterion density is in excellent agreement 
with MC data for strong and intermediate Coulombic couplings.
Although the first correction term is small relative to the leading one
for small distances between the parallel walls, its precise form is important
when calculating the pressure between the charged walls via the contact-value 
theorem at larger distances and specifying regions of the couplings
and of the walls distances where the pressure is attractive.

The crucial problem with the Wigner SC approach is that Wigner crystals become 
unstable at extremely large values of the coupling constant; the melting of 
the single-layer and double-layer Wigner structures to their
fluid phases is described in references \cite{Strandburg88} and
\cite{Goldoni96,Schweigert99}, respectively.
In spite of this taking the Wigner lattice as a reference provides an adequate
description of the fluid phase up to intermediate couplings.
The strong Coulomb repulsion of identical charges causes that their pair
correlation function almost vanishes at small distances.
The idea of a correlation hole was applied successfully in various ways 
to go beyond the PB theory 
\cite{Bakhshandeh11,Barbosa00,Forsman04,Nordholm84,Rouzina96}.
To adapt the quantitatively correct Wigner SC approach to the fluid phase, 
the Wigner structure was substituted by a correlation hole in
\cite{Samaj16a}.
In the case of one wall with counterions only, another correlation-hole theory
of the self-consistent nature \cite{Palaia18} leads to a modified 
PB integral equation which implies the exact density profiles in 
both WC and SC limits. 
In contrast to similar attempts to establish a universal theory
working well for any coupling \cite{Burak04,Santangelo06},
the density profile satisfies the contact-value theorem and provides
a crossover from a short-distance exponential to a large-distance 
algebraic PB decay from the charged wall via a large density plateau. 

The WC and SC analyses were done explicitly on the exactly solvable 1D gas 
of counterions \cite{Dean09}.
As concerns the 2D problem of one line-charged wall with counterions only, 
the density profile at $\Gamma=2$ was derived by Jancovici \cite{Jancovici84}.
The pressure for two parallel walls at distance $d$ was obtained 
in the symmetric and nonsymmetric cases in references \cite{Samaj11c} 
and \cite{Samaj14}, respectively.
In the case of like-charged walls, the pressure is always positive and
decays monotonously from infinity at $d\to 0$ to $0$ as $d\to\infty$.
Another type of exact results concerns the Manning condensation of counterions 
at the charged surface of the 3D cylinder \cite{Burak06,Naji06}.

In equilibrium statistical mechanics of fluid systems it is generally 
believed that, except for phase transitions, a few particles are able to
reproduce adequately statistical quantities of large systems \cite{Ma}.
The primary motivation for the present work is the absence of exact results 
for 2D one-component models with the coupling constants $\Gamma>2$ where one 
expects the counterintuitive phenomenon of the attraction between 
like-charged walls.
We consider the cylinder of circumference $W$ and finite length $d$ with 
the charged circle ends, the counterions are allowed to move freely on the 
cylinder surface; such a model is well defined also for finite numbers of 
particles $N$.
As is shown in this paper for the exactly solvable free-fermion coupling 
$\Gamma=2$, the results for the pressure as the function of $d$ for $N=8$ 
particles turn out to be very close to those for $N\to\infty$ particles.
This fact justifies the exact treatment of the couplings $\Gamma=4$ and $6$ 
up to $N=8$ particles by using the anticommuting-field formalism 
\cite{Samaj95} which can be done with modest computational efforts. 
It turns out that the attraction phenomenon of like-charged walls
is observed for these relatively small couplings.
As a by-product of the anticommuting-field formalism, we derive within 
the cylinder geometry the exact constraints (sum rules) for the particle 
one-body and two-body densities which have direct impact on characteristics 
of the long-range decay of two-body densities along the two walls in 
the pure 2D limit $W\to\infty$. 

The paper is organized as follows.
In Sect. \ref{Sec:2}, we review the general formalism for Coulomb systems 
confined to the surface of a cylinder and their mapping onto the 1D lattice 
model of interacting anticommuting fields for the coupling constant
$\Gamma=2\gamma$ with $\gamma$ a positive integer.
The exact cylinder sum rules for the particle one-body and two-body
densities are derived in Sect. \ref{Sec:3}.
The impact of these sum rules on the long-range decay of 2D two-body
densities along the two walls is explained in Sect. \ref{Sec:4}.
Sect. \ref{Sec:5} deals with the exactly solvable $\Gamma=2$ case.
The coupling constants $\Gamma=4$ and $6$ are treated for
a finite number of particles in Sect. \ref{Sec:6}. 
The concluding Sect. \ref{Sec:7} is a short recapitulation. 
  
\renewcommand{\theequation}{2.\arabic{equation}}
\setcounter{equation}{0}

\section{General formalism for cylinder geometry} \label{Sec:2}

\subsection{Cylinder geometry} \label{Sec:21}
\begin{figure}[tbh]
\begin{center}
\includegraphics[width=0.7\textwidth,clip]{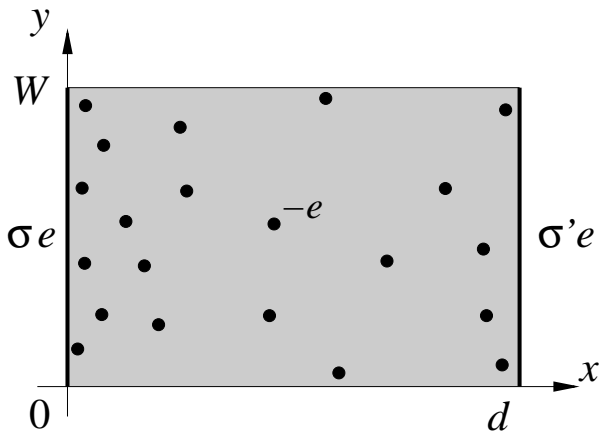}
\caption{The cylinder geometry with the periodic boundary conditions 
(period $W$) along the $y$-axis. 
Two parallel lines (circles) with the fixed charge densities $\sigma e$ 
and $\sigma' e$ are localized at the end points $x=0$ and $x=d$, respectively. 
Pointlike counterions of charge $-e$ are allowed to move freely between 
the two charged lines.}
\label{Fig:1}
\end{center}
\end{figure}

We consider the system of $N$ mobile pointlike particles with 
the elementary charge $-e$, confined to the surface of a cylinder of 
circumference $W$ and length $d$.
The surface of the cylinder can be represented equivalently as a 2D 
semiperiodic rectangle domain $\Lambda$ of points ${\bf r}=(x,y)$ with 
coordinates $x\in [0,d]$ (no restricting conditions at the end-points $x=0,d$) 
and $y\in [0,W]$ (periodic boundary conditions at $y=0,W$), 
see Fig. \ref{Fig:1}.
It is useful to introduce the complex coordinates $z=x+{\rm i}y$ and
$\bar{z}=x-{\rm i}y$.
There are the fixed uniform line charge densities $\sigma e$ and $\sigma' e$
($\sigma,\sigma'$ having dimension [length]$^{-1}$) along the $y$-axis at 
the end-points $x=0$ and $x=d$, respectively.
We restrict ourselves to the like-charged line segments (circles), 
i.e., without any loss of generality, $0\le\sigma'\le\sigma$.
Introducing the asymmetry parameter
\begin{equation}
\eta \equiv \frac{\sigma'}{\sigma} , \qquad \eta\in [0,1] ,
\end{equation}
the symmetric case $\sigma=\sigma'$ corresponds to $\eta=1$.
The overall electroneutrality condition is expressed as
\begin{equation}
N = (\sigma+\sigma') W .
\end{equation} 
The thermodynamic limit corresponds to the limits $N,W\to\infty$, 
keeping the ratio $N/W=\sigma+\sigma'$ fixed.
The system possesses the obvious exchange symmetry 
$\sigma\leftrightarrow\sigma'$ under the coordinate transformation $x\to d-x$. 
The dielectric constants of the walls $\varepsilon_W$ and of the medium 
the particles are immersed in $\varepsilon$ are considered to be the same, 
$\varepsilon_W=\varepsilon$, i.e., there are no image charges. 

The Coulomb potential $\phi$ at a spatial position ${\bf r}\in\Lambda$, induced
by a unit charge at the origin ${\bf 0}$, is defined as the solution of 
the 2D Poisson equation 
$\Delta \phi({\bf r}) = -2\pi\delta({\bf r})/\varepsilon$, under
the periodicity requirement along the $y$-axis with period $W$.
Considering the potential as a Fourier series in $y$, it is obtained in
the form \cite{Choquard81}  
\begin{equation} \label{periodicCoulomb}
\phi({\bf r}) = \frac{1}{\varepsilon W} \sum_{k_y} \int_{-\infty}^{\infty} 
{\rm d}k_x \frac{1}{k_x^2+k_y^2} {\rm e}^{{\rm i}(k_x x + k_y y)} , 
\qquad k_y\in \frac{2\pi n}{W}
\end{equation}
with $n=0,\pm 1,\ldots$ being any integer.
It is seen that also in the mixed discrete-continuous Fourier representation
of the Coulomb potential has the characteristic $1/k^2$ form.
After integration over $k_x$ and summation over $k_y$, the periodic
Coulomb potential (\ref{periodicCoulomb}) takes the form
\begin{eqnarray}
\phi({\bf r}) & = & - \frac{1}{\varepsilon}
\ln \left\vert 2 \sinh\left( \frac{\pi z}{W} \right)
\right\vert \nonumber \\ & = & - \frac{1}{2\varepsilon}
\ln \left[ 2 \cosh\left( \frac{2\pi x}{W} \right) - 
2 \cos\left( \frac{2\pi y}{W} \right) \right] .
\end{eqnarray} 
For small distances $r\ll W$, this potential reduces to the 2D
Coulomb one $-(1/\varepsilon)\ln(2\pi r/W)$.
At large distances along the cylinder $x\gg W$, this potential behaves like
the 1D Coulomb one $-\pi \vert x\vert/(\varepsilon W)$.
For the calculation of the Coulomb interaction between charge line densities
and particles, the following formula is important: 
\begin{equation} \label{formula1}
\int_0^W {\rm d} y\, \phi({\bf r}) = - \frac{\pi}{\varepsilon} x .
\end{equation}

According to the analysis made in \cite{Samaj14}, the Coulomb energy of 
$N$ particles at spatial positions $\{ {\bf r}_1, \cdots, {\bf r}_N \}$ plus 
the fixed line charge densities $\sigma e$ and $\sigma' e$ consists of 
the self and mutual interactions of the line charge densities 
$E_{ll}=-\pi\sigma\sigma' W d e^2/\varepsilon$, of the interaction of 
particles with line charge densities 
$E_{pl} = \sum_{j=1}^N \pi(\sigma-\sigma') x_j e^2/\varepsilon + 
N\pi\sigma' d e^2/\varepsilon$ and the pair interactions of the particles 
$E_{pp} = \sum_{(j<k)=1}^N e^2 \phi(\vert {\bf r}_j-{\bf r}_k\vert)$. 
At inverse temperature $\beta=1/(k_{\rm B}T)$, the Boltzmann factor of
the total energy $E_N=E_{ll}+E_{pl}+E_{pp}$ reads as 
\begin{equation}
{\rm e}^{-\beta E_N(\{{\bf r}\})} = {\rm e}^{-\pi\Gamma(\sigma')^2 W d}
\prod_{j=1}^N {\rm e}^{-\beta v(x_j)} 
\prod_{(j<k)=1}^N \left\vert 2 \sinh\frac{\pi(z_j-z_k)}{W} \right\vert^{\Gamma} ,
\end{equation}
where $v(x)$ is the one-body potential energy given by
\begin{equation} \label{vx}
\beta v(x) = \pi\Gamma (\sigma-\sigma') x
\end{equation} 
and $\Gamma=\beta e^2/\varepsilon$ is the coupling constant.
Within the canonical ensemble, the partition function is defined as
\begin{equation} \label{partition}
Z_N(\gamma) = \frac{1}{N!} \int_{\Lambda} \frac{{\rm d}{\bf r}_1}{\lambda^2} 
\cdots \int_{\Lambda} \frac{{\rm d}{\bf r}_N}{\lambda^2} 
{\rm e}^{-\beta E_N(\{{\bf r}\})} ,
\end{equation}
where $\lambda$ is the thermal de Broglie wavelength. 

There exist two possible representations of the partition function.

Firstly, applying the formula
\begin{equation} \label{formula2}
\left\vert 2 \sinh \frac{\pi (z-z')}{W} \right\vert
= {\rm e}^{\frac{\pi}{W}(x+x')} \left\vert
{\rm e}^{-\frac{2\pi}{W}z} - {\rm e}^{-\frac{2\pi}{W}z'} \right\vert
\end{equation}
to each two-particle interaction Boltzmann factor the partition function 
can be reexpressed as
\begin{equation} \label{partf}
Z_N(\gamma) = \left( \frac{W^2}{4\pi\lambda^2} \right)^N 
\exp\left[ -\pi\Gamma(\sigma')^2 W d \right] Q_N(\gamma) ,
\end{equation}
where
\begin{equation} \label{partgen}
Q_N(\gamma) = \frac{1}{N!} \int_{\Lambda} \prod_{j=1}^N \left[ {\rm d}^2 z_j\, 
w_{\rm ren}({\bf r}_j) \right]
\prod_{j<k} \left\vert {\rm e}^{-\frac{2\pi}{W}z_j} - {\rm e}^{-\frac{2\pi}{W}z_k}
\right\vert^{\Gamma} 
\end{equation}
with the renormalized one-body Boltzmann factor
$w_{\rm ren}({\bf r})\equiv w_{\rm ren}(x)$ given by
\begin{equation} \label{onebody}
w_{\rm ren}(x) = \frac{4\pi}{W^2} \exp\left[ -\beta v(x)
+ \frac{\pi\Gamma}{W} (N-1) x \right] .
\end{equation}

The second representation of the partition function follows from 
another version of the formula (\ref{formula2}):
\begin{equation} \label{formula3}
\left\vert 2 \sinh \frac{\pi (z-z')}{W} \right\vert
= {\rm e}^{-\frac{\pi}{W}(x+x')} \left\vert
{\rm e}^{\frac{2\pi}{W}z} - {\rm e}^{\frac{2\pi}{W}z'} \right\vert
\end{equation}
Then the partition function is still given by (\ref{partf}) where
\begin{equation} \label{partgen2}
Q_N(\gamma) = \frac{1}{N!} \int_{\Lambda} \prod_{j=1}^N \left[ {\rm d}^2 z_j\, 
w_{\rm ren}({\bf r}_j) \right]
\prod_{j<k} \left\vert {\rm e}^{\frac{2\pi}{W}z_j} - {\rm e}^{\frac{2\pi}{W}z_k}
\right\vert^{\Gamma} 
\end{equation}
with the renormalized one-body Boltzmann factor
\begin{equation} \label{onebody2}
w_{\rm ren}(x) = \frac{4\pi}{W^2} \exp\left[ -\beta v(x)
- \frac{\pi\Gamma}{W} (N-1) x \right] .
\end{equation}

In what follows, we shall use mainly the first representation 
(\ref{formula2})-(\ref{onebody}). 
The free energy $F_N$, defined by $-\beta F_N = \ln Z_N$, is expressible 
in both cases as
\begin{equation} \label{free}
-\beta F_N(\gamma) = N \ln \left( \frac{W^2}{4\pi\lambda^2} \right) 
-\pi\Gamma(\sigma')^2 W d + \ln Q_N(\gamma) .
\end{equation}

The particle density at point ${\bf r}\in \Lambda$ is given by 
\begin{equation}
n({\bf r}) = \left\langle \hat{n}({\bf r}) \right\rangle , 
\qquad \hat{n}({\bf r}) = \sum_{j=1}^N \delta({\bf r}-{\bf r}_j) , 
\end{equation}
where $\langle \cdots \rangle$ denotes the statistical average over
the canonical ensemble and $\hat{n}({\bf r})$ is the microscopic particle
number density.  
The particle density can be obtained in the standard way as the functional 
derivative
\begin{equation}
n({\bf r}) = w_{\rm ren}({\bf r}) \frac{1}{Q_N} 
\frac{\delta Q_N}{\delta w_{\rm ren}({\bf r})} .
\end{equation}
Since the one-body potential (\ref{vx}) depends on the $x$-coordinate only 
and due to the cylinder geometry, it holds that $n({\bf r})\equiv n(x)$. 

The two-body density
\begin{equation}
n^{(2)}({\bf r},{\bf r'}) = \left\langle \sum_{(j\ne k)=1}^N 
\delta({\bf r}-{\bf r}_j) \delta({\bf r}'-{\bf r}_k) \right\rangle 
\end{equation}
can be calculated as
\begin{equation}
n^{(2)}({\bf r},{\bf r}') = w_{\rm ren}({\bf r}) w_{\rm ren}({\bf r}') 
\frac{1}{Q_N} \frac{\delta^2 Q_N}{\delta w_{\rm ren}({\bf r}) 
\delta w_{\rm ren}({\bf r}')} .
\end{equation}
The corresponding (truncated) Ursell function $U$ and the density structure 
function $S$ are defined by
\begin{eqnarray} 
U({\bf r},{\bf r}') & = & n^{(2)}({\bf r},{\bf r'}) - n({\bf r}) n({\bf r}') ,  
\label{Ursell} \\
S({\bf r},{\bf r}') & = & \left\langle \hat{n}({\bf r}) \hat{n}({\bf r}')
\right\rangle - n({\bf r}) n({\bf r}') = U({\bf r},{\bf r}') 
+ n({\bf r}) \delta({\bf r}-{\bf r}') , \label{structure}
\end{eqnarray}
respectively.
Due to the cylinder geometry, the two-point functions $n^{(2)}$, $U$ and
$S$ depend on $x$, $x'$ and $\vert y-y'\vert$.

\subsection{Mapping onto the 1D lattice anticommuting-field theory} 
\label{Sec:22}
For $\Gamma=2\gamma$ ($\gamma=1,2,3,\ldots$ a positive integer), 
the technique of anticommuting variables \cite{Samaj95,Samaj04a} allows us to 
express $Q_N$ (\ref{partgen}) as an integral over Grassman variables; 
for the cylinder geometry the mapping is established in \cite{Samaj04b,Samaj14}.
Let us consider a discrete chain of $N$ sites $j=0,1,\ldots,N-1$.
At each site $j$, there is $\gamma$ variables of type $\{ \xi_j^{(\alpha)}\}$
and $\gamma$ variables of type $\{ \psi_j^{(\alpha)}\}$ 
$(\alpha=1,\ldots,\gamma)$, all variables anticommute with each other. 
The multi-dimensional integral of the form (\ref{partgen}) can be expressed as
the integral over anticommuting variables:
\begin{equation} \label{antipart}
Q_N(\gamma) = \int {\cal D}\psi {\cal D}\xi\, {\rm e}^{S(\Xi,\Psi)} , 
\qquad S(\Xi,\Psi) = \sum_{j=0}^{\gamma(N-1)} \Xi_j w_j \Psi_j .
\end{equation}
Here, ${\cal D}\psi {\cal D}\xi \equiv \prod_{j=0}^{N-1} {\rm d}\psi_j^{(\gamma)}
\cdots {\rm d}\psi_j^{(1)} {\rm d}\xi_j^{(\gamma)} \cdots {\rm d}\xi_j^{(1)}$
and the action $S(\Xi,\Psi)$ involves pair interactions of composite
operators
\begin{equation} \label{composite}
\Xi_j = \sum_{j_1,\ldots,j_{\gamma}=0\atop (j_1+\cdots+j_{\gamma}=j)}^{N-1}
\xi_{j_1}^{(1)} \cdots \xi_{j_{\gamma}}^{(\gamma)} , \qquad
\Psi_j = \sum_{j_1,\ldots,j_{\gamma}=0\atop (j_1+\cdots+j_{\gamma}=j)}^{N-1}
\psi_{j_1}^{(1)} \cdots \psi_{j_{\gamma}}^{(\gamma)} ,
\end{equation} 
i.e. the products of $\gamma$ anticommuting variables of one type with 
the prescribed sum of site indices.
The interaction strengths $w_j$ $[j=0,1,\ldots,\gamma(N-1)]$ are given by
\begin{eqnarray} 
w_j & = & W \int_0^d {\rm d} x\, w_{\rm ren}(x) 
\exp\left( - \frac{4\pi}{W} j x \right) \nonumber \\ 
& = & \frac{1-\exp\left[ - \frac{4\pi d}{W}\left(j-\gamma W\sigma'
+\frac{\gamma}{2}\right)\right]}{j-\gamma W\sigma'+\frac{\gamma}{2}} .
\label{wj}
\end{eqnarray}

The main advantage of the present formalism is that 
the one-body and two-body particle densities are expressible
explicitly in terms of averages over the anticommuting variables
\begin{equation}
\langle \cdots\rangle \equiv \frac{1}{Q_N(\gamma)}
\int {\cal D}\psi {\cal D}\xi\, {\rm e}^{S(\Xi,\Psi)} \cdots
\end{equation}
of certain products of composite operators.
Namely, the particle density at $x$ is given by 
\begin{equation} \label{antione}
n(x) = w_{\rm ren}(x) \sum_{j=0}^{\gamma(N-1)} \langle \Xi_j \Psi_j \rangle
\exp\left( - \frac{4\pi}{W} j x \right) ,
\end{equation}
the two-body density between points ${\bf r}_1=(z_1,\bar{z}_1)$ and
${\bf r}_2=(z_2,\bar{z}_2)$ is expressible as
\begin{eqnarray} 
n^{(2)}(z_1,\bar{z}_1;z_2,\bar{z}_2) & = & w_{\rm ren}(x_1) w_{\rm ren}(x_2) 
\sum_{j_1,k_1,j_2,k_2=0\atop (j_1+j_2=k_1+k_2)}^{\gamma(N-1)} 
\langle \Xi_{j_1} \Psi_{k_1} \Xi_{j_2} \Psi_{k_2} \rangle \nonumber \\
& & \times \exp\left[ - \frac{2\pi}{W} 
\left( j_1 z_1 + k_1 \bar{z}_1 + j_2 z_2 + k_2 \bar{z}_2 \right) \right] .
\label{antitwo}
\end{eqnarray}

As a trivial application of the formalism, we derive the basic formula
of the contact-value theorem.
The pressure $P_N$ is the force between the charged circles, calculated
per unit length of one of the circles:
\begin{equation} \label{PNN}
\beta P_N = \frac{\partial}{\partial d} \left( \frac{-\beta F_N}{W} \right) .
\end{equation}
The two circles repel (attract) one another if the pressure is positive
(negative).
Using the formula (\ref{free}) for $F_N$, the anticommuting representation 
(\ref{antipart}) of $Q_N$ and the relation
\begin{equation}
\frac{\partial w_j}{\partial d} = W w_{\rm ren}(d) 
\exp\left( - \frac{4\pi}{W} j d \right) , 
\end{equation}
we arrive at the relationships given by the contact-value theorem 
\begin{equation} \label{PN}
\beta P_N = n(d) - 2\pi\gamma \left( \sigma'\right)^2 =
n(0) - 2\pi\gamma \sigma^2 ,
\end{equation}
where the second equality follows directly from the invariance of the pressure
with respect to the exchange symmetry $\sigma\leftrightarrow\sigma'$ under 
the coordinate transformation $x\to d-x$. 
Note that if $\sigma'=0$ it holds that $\beta P_N = n(d)$ and since the
particle density is a positive quantity the pressure cannot be negative.  
We shall often use the notation
\begin{equation} \label{Ptilde}
\tilde{P} \equiv \frac{\beta P}{2\pi\gamma\sigma^2}
= \frac{n(0)}{2\pi\gamma\sigma^2} - 1
\end{equation}
for the dimensionless pressure.
Since the particle density $n(0)$ is positive, it holds that $\tilde{P}\ge -1$;
the value $\tilde{P}=-1$ corresponds to the strongest possible attraction 
between the two charged walls.

\renewcommand{\theequation}{3.\arabic{equation}}
\setcounter{equation}{0}

\section{Sum rules for the cylinder geometry} \label{Sec:3}
As was shown in \cite{Samaj00}, there exist specific linear
transformations of anticommuting variables which keep the composite form
of the composite operators (\ref{composite}).
Most of transformations consist in a simple rescaling of one or all 
anticommuting components of a given $\xi$ or $\psi$ type, however, 
there is one nontrivial transformation which mixes all anticommuting-field 
components of a given type.
Each transformation leads to the exact constraints (sum rules) for the 
correlation functions of the composite operators whose forms are universal in 
the sense that they {\em do not depend} on the particular form of 
the interaction strengths $\{ w_j \}_{j=0}^{\gamma(N-1)}$.
These sum rules can be used to derive integral/differential equations for
the one- and two-body densities whose forms {\em depend} on 
the particular geometry of the Coulomb problem.
The one-body Boltzmann factor $w_{\rm ren}(x)$ will be considered in the
general form (\ref{onebody}) with $\beta v(x)$ given by (\ref{vx}) for
the present case of counterions in the potential of charged lines 
with density $\sigma e$ at $x=0$ and $\sigma' e$ at $x=d$.

\subsection{Scaling transformations}
$\bullet$ Rescaling by a constant $\mu$ one of the anticommuting field 
components, say
\begin{equation} \label{trans1}
\xi_j^{(1)} \to \mu \xi_j^{(1)} , \qquad j=0,1,\ldots,N-1 ,
\end{equation}
the composite operators $\Xi_j$ get the same factor $\mu$ and the action
in (\ref{antipart}) transforms itself as $S(\Xi,\Psi)\to \mu S(\Xi,\Psi)$.
The Jacobian of the transformation (\ref{trans1}) equals to $\mu^N$. 

Under the transformation (\ref{trans1}), the quantity $Q_N$ (\ref{antipart})
takes the form 
\begin{equation}
Q_N = \mu^{-N} \int {\cal D}\psi {\cal D}\xi
\exp\left( \mu \sum_{j=0}^{\gamma(N-1)} \Xi_j w_j \Psi_j \right) .
\end{equation}
Since $Q_N$ does not depend on $\mu$, it holds that 
$\partial \ln Q_N/\partial\mu\vert_{\mu=1} = 0$ or, equivalently,
\begin{equation} \label{sr11}
\sum_{j=0}^{\gamma(N-1)} w_j \langle \Xi_j \Psi_j \rangle = N .
\end{equation}
Consequently,
\begin{equation} \label{ie11}
W \int_0^d {\rm d}x\, n(x) = N ,
\end{equation}
where we have substituted $n(x)$ from (\ref{antione}) and used the definition
of the interaction strength $w_j$ (\ref{wj}).
This equation provides the trivial information that there are $N$ particles 
inside the cylinder domain $\Lambda$.

Applying the transformation (\ref{trans1}) to the quantity
$Q_N \langle\Xi_j\Psi_j\rangle$, one gets 
\begin{equation}
Q_N \langle \Xi_j \Psi_j \rangle = \mu^{-N+1} \int {\cal D}\psi {\cal D}\xi 
\Xi_j \Psi_j \exp\left( \mu \sum_{k=0}^{\gamma(N-1)} \Xi_k w_k \Psi_k \right) . 
\end{equation}
The equality
$\partial (Q_N\langle\Xi_j\Psi_j\rangle)/\partial\mu\vert_{\mu=1} = 0$ 
implies that
\begin{equation} \label{sr12}
\sum_{k=0}^{\gamma(N-1)} w_k \langle \Xi_j \Psi_j \Xi_k \Psi_k \rangle 
= (N-1) \langle \Xi_j \Psi_j \rangle .
\end{equation}
This sum rule can be transformed into an integral equation by considering
\begin{eqnarray} 
\int_0^d {\rm d}x' \int_0^W {\rm d}y\, n^{(2)}(x,x';y) & = &
w_{\rm ren}(x) \sum_{j,k=0}^{\gamma(N-1)} w_k 
\langle \Xi_j \Psi_j \Xi_k \Psi_k \rangle \nonumber \\ & & \times
\exp\left( -\frac{4\pi}{W} jx\right) \label{eq1} ,
\end{eqnarray} 
where we have inserted the representation (\ref{antitwo}) of the two-body
density and applied the orthogonality relation
\begin{equation}
\int_0^W {\rm d}y\, \exp\left[ - \frac{2\pi}{W} {\rm i} (j-k) y \right]
= W \delta_{j,k} .
\end{equation}
Considering the sum rule (\ref{sr12}) for the sum over $k$ on the rhs of
(\ref{eq1}), the rhs becomes equal to $(N-1) n(x)$ and
\begin{equation} \label{Un}
\int_0^d {\rm d}x' \int_0^W {\rm d}y\, U(x,x';y) = (N-1) n(x) - N n(x) 
= -n(x) . 
\end{equation}
Consequently,
\begin{equation} \label{zerothmoment}
\int_0^d {\rm d}x' \int_0^W {\rm d}y\, S(x,x';y) = 0 . 
\end{equation}
This relation represents a generalization of the zeroth-moment
Stillinger-Lovett condition \cite{Stillinger68a,Stillinger68b}
to the cylinder geometry.

$\bullet$ Let us rescale all anticommuting field $\xi$-components as follows
\begin{equation} \label{trans2}
\xi_j^{(\alpha)} \to \lambda^j \xi_j^{(\alpha)} , \qquad j=0,1,\ldots,N-1 ,
\qquad \alpha=1,\ldots,\gamma . 
\end{equation}
The composite operators $\Xi_j$ acquire the factor $\lambda^j$ 
and the action in (\ref{antipart}) transforms itself as 
$S(\Xi,\Psi)\to \sum_{j=0}^{\gamma(N-1)} \lambda^j \Xi_j w_j \Psi_j$.
The Jacobian of the transformation (\ref{trans2}) equals to 
$\lambda^{\gamma N(N-1)/2}$. 

Under the transformation (\ref{trans2}), the quantity $Q_N$ (\ref{antipart})
is rewritten as
\begin{equation}
Q_N = \lambda^{-\gamma N(N-1)/2} \int {\cal D}\psi {\cal D}\xi
\exp\left( \sum_{j=0}^{\gamma(N-1)} \lambda^j \Xi_j w_j \Psi_j \right) .
\end{equation}
The requirement $\partial \ln Q_N/\partial\lambda\vert_{\lambda=1} = 0$ 
is equivalent to the sum rule
\begin{equation} \label{sr21}
\sum_{j=0}^{\gamma(N-1)} j w_j \langle \Xi_j \Psi_j \rangle 
= \frac{1}{2}\gamma N (N-1) .
\end{equation}
To make use of this relation, we consider the integral
\begin{eqnarray} 
W \int_0^d {\rm d}x\, w_{\rm ren}(x) \frac{\partial}{\partial x}
\left[ \frac{n(x)}{w_{\rm ren}(x)} \right] & = & \sum_{j=0}^{\gamma(N-1)}
\langle \Xi_j\Psi_j \rangle W \int_0^d {\rm d}x\, w_{\rm ren}(x) 
\nonumber \\ & & \times \left( - \frac{4\pi j}{W} \right) 
\exp\left( -\frac{4\pi}{W} jx\right) . \label{ie21}
\end{eqnarray}
With regard to the definition of the interaction strengths 
$\{ w_j \}$ (\ref{wj}), the rhs of this equation equals to $-4\pi/W$ times 
the lhs of Eq. (\ref{sr21}), so that
\begin{equation}
\int_0^d {\rm d}x\, \frac{\partial}{\partial x} n(x) 
- \int_0^d {\rm d}x\, n(x) \frac{\partial}{\partial x} \ln w_{\rm ren}(x)
= - \frac{2\pi}{W^2} \gamma N (N-1) .
\end{equation} 
Since
\begin{equation} \label{partialw}
\frac{\partial}{\partial x} \ln w_{\rm ren}(x) = - \frac{\partial}{\partial x}
\left[ \beta v(x) \right] + \frac{2\pi\gamma}{W} (N-1) ,
\end{equation}
we finally end up with the relation
\begin{equation}
n(d) - n(0) = - \int_0^d {\rm d}x\, n(x) \frac{\partial}{\partial x}
\left[ \beta v(x) \right] .
\end{equation}
For our system of counterions with $\beta v(x)$ given by (\ref{vx}),
one obtains the equality
\begin{equation}
n(d) - 2\pi\gamma \left( \sigma'\right)^2 =
n(0) - 2\pi\gamma \sigma^2
\end{equation}
of two possible representations of the pressure (\ref{PN}).

The application of the transformation (\ref{trans2}) to 
$Q_N \langle\Xi_j\Psi_j\rangle$ results in
\begin{equation}
Q_N \langle \Xi_j \Psi_j \rangle = \lambda^{-\gamma N(N-1)/2+j} 
\int {\cal D}\psi {\cal D}\xi\, \Xi_j \Psi_j \exp\left( 
\sum_{k=0}^{\gamma(N-1)} \lambda^k \Xi_k w_k \Psi_k \right) . 
\end{equation}
The equality
$\partial (Q_N\langle\Xi_j\Psi_j\rangle)/\partial\lambda\vert_{\lambda=1} = 0$ 
leads to
\begin{equation} \label{sr22}
\sum_{k=0}^{\gamma(N-1)} k w_k \langle \Xi_j \Psi_j \Xi_k \Psi_k \rangle 
= \left[ \frac{1}{2}\gamma N (N-1) - j \right] \langle \Xi_j \Psi_j \rangle .
\end{equation}
This equation can be rewritten with the aid of the sum rule (\ref{sr21}) 
as follows 
\begin{equation} \label{sr22prime}
\sum_{k=0}^{\gamma(N-1)} k w_k \langle \Xi_j \Psi_j \Xi_k \Psi_k \rangle^{\rm T} 
= - j \langle \Xi_j \Psi_j \rangle ,
\end{equation}
where the truncated correlators 
$\langle \Xi_j \Psi_j \Xi_k \Psi_k \rangle^{\rm T} \equiv
\langle \Xi_j \Psi_j \Xi_k \Psi_k \rangle - \langle \Xi_j \Psi_j \rangle
\langle \Xi_k \Psi_k \rangle$.
Let us consider the integral
\begin{eqnarray} 
\int_0^d {\rm d}x' \int_0^W {\rm d}y\,  w_{\rm ren}(x') 
\frac{\partial}{\partial x'} \left[ \frac{U(x,x';y)}{w_{\rm ren}(x')} \right] 
& & \nonumber \\ 
= w_{\rm ren}(x) \sum_{j,k=0}^{\gamma(N-1)} w_k
\langle \Xi_j\Psi_j \Xi_k \Psi_k \rangle^{\rm T} \left( - \frac{4\pi k}{W} \right)
\exp\left( -\frac{4\pi}{W} jx\right) . & & \label{ie22}
\end{eqnarray}  
With regard to the sum rule (\ref{sr22prime}), the rhs of this equation
is written as
\begin{equation}
- w_{\rm ren}(x) \frac{\partial}{\partial x}
\left[ \frac{n(x)}{w_{\rm ren}(x)} \right] = - \frac{\partial n(x)}{\partial x}
+ n(x) \frac{\partial}{\partial x} \left[ \ln w_{\rm ren}(x) \right] . 
\end{equation}
The lhs of Eq. (\ref{ie22}) can be expressed as
\begin{equation}
\int_0^d {\rm d}x' \int_0^W {\rm d}y\, \left\{ \frac{\partial}{\partial x'} 
U(x,x';y) - U(x,x';y) \frac{\partial}{\partial x'} 
\left[ \ln w_{\rm ren}(x') \right] \right\} .
\end{equation}
Using the relations (\ref{partialw}) and (\ref{Un}), we end up with
\begin{eqnarray}
\int_0^W {\rm d} y \left[ U(x,d;y) - U(x,0;y) \right] & & \nonumber \\
+ \int_0^d {\rm d}x' \int_0^W {\rm d}y\, S(x,x';y)
\frac{\partial}{\partial x'} \left[ \beta v(x') \right] 
& = & - \frac{\partial n(x)}{\partial x} .
\end{eqnarray} 
For $\beta v(x)$ given by (\ref{vx}), with regard to (\ref{zerothmoment})
this equation simplifies itself to the one
\begin{equation} \label{Wertheim}
\frac{\partial n(x)}{\partial x} =
\int_0^W {\rm d} y \left[ U(x,0;y) - U(x,d;y) \right] 
\end{equation} 
which is a generalization of the 2D Wertheim-Lovett-Mou-Buff
(WLMB) equation \cite{Lovett76,Wertheim76} to the surface of cylinder 
constrained by two charged lines.

\subsection{Transformation mixing all anticommuting components}
It was shown in \cite{Samaj00} that there exists a nontrivial
transformation of anticommuting variables, say $\xi'$s,
\begin{equation} \label{trans3}
\xi_j^{(\alpha)}(t) = \sum_{k=j}^{N-1} {k\choose j} t^{k-j} \xi_k^{(\alpha)} ,
\quad j=0,1,\ldots,N-1 , \quad \alpha=1,\ldots,\gamma ,  
\end{equation}
which keeps the composite form of the transformed composite operators:
\begin{equation}
\Xi_j(t) = \sum_{k=j}^{\gamma(N-1)} {k\choose j} t^{k-j} 
\Xi_k , \qquad j=0,1,\ldots,\gamma(N-1) .  
\end{equation}
Here, $t$ is a free parameter; the case $t=0$ corresponds to the
identity mapping.
The Jacobian of the transformation equals to 1.

Under the transformation (\ref{trans3}), $Q_N$ (\ref{antipart})
takes the form
\begin{eqnarray}
Q_N & = & \int {\cal D}\psi {\cal D}\xi
\exp\left[ \sum_{j=0}^{\gamma(N-1)} \Xi_j(t) w_j \Psi_j \right] \nonumber \\ 
& = & \int {\cal D}\psi {\cal D}\xi
\exp\left\{ \sum_{j=0}^{\gamma(N-1)} \left[ \Xi_j + t (j+1) \Xi_{j+1} 
+ O(t^2) \right] w_j \Psi_j \right\} . \phantom{aa} 
\end{eqnarray}
The condition $\partial \ln Q_N/\partial t\vert_{t=0} = 0$ implies the sum rule
\begin{equation} \label{sr31}
\sum_{j=0}^{\gamma(N-1)-1} (j+1) w_j \langle \Xi_{j+1} \Psi_j \rangle = 0 .
\end{equation}
This sum rule is trivial because the diagonalized action (\ref{antipart})
implies that every correlator $\langle \Xi_{j+1} \Psi_j \rangle = 0$.

Applying the transformation (\ref{trans3}) to 
$Q_N \langle\Xi_{j-1}\Psi_j\rangle$ implies 
\begin{eqnarray}
Q_N \langle \Xi_{j-1} \Psi_j \rangle & = & \int {\cal D}\psi {\cal D}\xi\, 
\Xi_{j-1}(t) \Psi_j \exp\left[ \sum_{k=0}^{\gamma(N-1)} \Xi_k(t) w_k \Psi_k \right]  
\nonumber \\ & = & \int {\cal D}\psi {\cal D}\xi 
\left[ \Xi_{j-1} + t j \Xi_j + O(t^2) \right] \Psi_j \nonumber \\ & & 
\times \exp\left\{ \sum_{k=0}^{\gamma(N-1)} 
\left[ \Xi_k + t (k+1) \Xi_{k+1} + O(t^2) \right]  w_k \Psi_k \right\} . 
\phantom{aaaaa}
\end{eqnarray}
The requirement
$\partial (Q_N\langle\Xi_j\Psi_j\rangle)/\partial t\vert_{t=0} = 0$ leads to
\begin{equation} \label{sr32}
\sum_{k=0}^{\gamma(N-1)-1} (k+1) w_k \langle \Xi_{j-1} \Psi_j \Xi_{k+1} \Psi_k \rangle 
= - j\langle \Xi_j \Psi_j \rangle .
\end{equation}
To make use of this sum rule, let us consider the integral
\begin{eqnarray}
\int_0^W {\rm d}y\, {\rm e}^{-{\rm i}\frac{2\pi}{W}y} n^{(2)}(x,x';y)
& = & w_{\rm ren}(x) w_{\rm ren}(x') W \sum_{j,k} 
\langle \Xi_{j-1} \Psi_j \Xi_{k+1} \Psi_k \rangle \nonumber \\ & & \times
{\rm e}^{-\frac{2\pi}{W}(2j-1)x} {\rm e}^{-\frac{2\pi}{W}(2k+1)x'} . 
\end{eqnarray}
Note that $n^{(2)}(x,x';y)$ can be substituted by $U(x,x';y)$ in this relation
and since $U(x,x';y)=U(x,x';-y)$ only the real part of
${\rm e}^{-{\rm i}\frac{2\pi}{W}y} = \cos\left(\frac{2\pi}{W}y\right)
- {\rm i}\sin\left(\frac{2\pi}{W}y\right)$ survives. 
Consequently,
\begin{eqnarray} 
\int_0^d {\rm d}x' \int_0^W {\rm d}y\, \cos\left(\frac{2\pi}{W}y\right)
w_{\rm ren}(x') {\rm e}^{\frac{4\pi}{W}x'} \frac{\partial}{\partial x'} 
\left[ \frac{U(x,x';y) {\rm e}^{-\frac{2\pi}{W}x'}}{w_{\rm ren}(x')} \right] 
& & \nonumber \\ 
= w_{\rm ren}(x) \left( - \frac{4\pi}{W} \right)
\sum_{j,k} (k+1) w_k \langle \Xi_{j-1} \Psi_j \Xi_{k+1} \Psi_k \rangle 
{\rm e}^{-\frac{2\pi}{W}(2j-1)x} . \label{ie32}
\end{eqnarray}  
Applying the sum rule (\ref{sr32}), the rhs of this equation can be 
expressed as
\begin{eqnarray}
- w_{\rm ren}(x) {\rm e}^{\frac{2\pi}{W}x} \frac{\partial}{\partial x}
\left[ \frac{n(x)}{w_{\rm ren}(x)} \right] & = & {\rm e}^{\frac{2\pi}{W}x} 
\left\{ - \frac{\partial n(x)}{\partial x}
+ n(x) \frac{\partial}{\partial x} \left[ \ln w_{\rm ren}(x) \right] \right\} .
\nonumber \\ & & 
\end{eqnarray}
After simple algebra we finally arrive at
\begin{eqnarray} 
\int_0^W {\rm d}y\, \cos\left(\frac{2\pi}{W}y\right) 
\left[ {\rm e}^{\frac{2\pi}{W}d} U(x,d;y) - U(x,0;y) \right] 
+ \int_0^d {\rm d}x' \int_0^W {\rm d}y\, & & \nonumber \\ 
\times \cos\left(\frac{2\pi}{W}y\right)
{\rm e}^{\frac{2\pi}{W}x'} S(x,x';y) \left\{ \frac{\partial}{\partial x'}
\left[ \beta v(x') \right] - \frac{\pi}{W} \left[ \Gamma(N-1) + 4 \right]
\right\} & & \nonumber \\ = - {\rm e}^{\frac{2\pi}{W}x} 
\left[ \frac{\partial n(x)}{\partial x} + \frac{4\pi}{W} n(x) \right] .
\label{result1}
\end{eqnarray}  
For the one-body potential (\ref{vx}) it holds that
\begin{equation} \label{dodatok1}
\frac{\partial}{\partial x'} \left[ \beta v(x') \right] 
- \frac{\pi}{W} \left[ \Gamma(N-1) + 4 \right] 
= - \left[ 2\pi \Gamma \sigma' + \frac{\pi}{W} (4-\Gamma) \right] .
\end{equation}

Another version of the above sum rule can be derived by using the alternative 
representation of the Coulomb system on the cylinder surface given by
Eqs. (\ref{formula3})-(\ref{onebody2}), with the renormalized 
one-body Boltzmann factor $w_{\rm ren}(x)$ defined by (\ref{onebody2}).
Within this representation, the interaction strengths $w_j$ 
$[j=0,1,\ldots,\gamma(N-1)]$ are given by
\begin{equation} 
w_j = W \int_0^d {\rm d} x\, w_{\rm ren}(x) 
\exp\left( \frac{4\pi}{W} j x \right) ,
\end{equation}
the particle density by 
\begin{equation} \label{antioneprime}
n(x) = w_{\rm ren}(x) \sum_{j=0}^{\gamma(N-1)} \langle \Xi_j \Psi_j \rangle
\exp\left( \frac{4\pi}{W} j x \right) 
\end{equation}
and the two-body density by
\begin{eqnarray} 
n^{(2)}(z_1,\bar{z}_1;z_2,\bar{z}_2) & = & w_{\rm ren}(x_1) w_{\rm ren}(x_2) 
\sum_{j_1,k_1,j_2,k_2=0\atop (j_1+j_2=k_1+k_2)}^{\gamma(N-1)} 
\langle \Xi_{j_1} \Psi_{k_1} \Xi_{j_2} \Psi_{k_2} \rangle \nonumber \\
& & \times \exp\left[ \frac{2\pi}{W} 
\left( j_1 z_1 + k_1 \bar{z}_1 + j_2 z_2 + k_2 \bar{z}_2 \right) \right] .
\label{antitwoprime}
\end{eqnarray}
The counterpart of the relation (\ref{ie32}) reads as
\begin{eqnarray} 
\int_0^d {\rm d}x' \int_0^W {\rm d}y\, \cos\left(\frac{2\pi}{W}y\right)
w_{\rm ren}(x') {\rm e}^{-\frac{4\pi}{W}x'} \frac{\partial}{\partial x'} 
\left[ \frac{U(x,x';y) {\rm e}^{\frac{2\pi}{W}x'}}{w_{\rm ren}(x')} \right] 
& & \nonumber \\ 
= w_{\rm ren}(x) \left( \frac{4\pi}{W} \right)
\sum_{j,k} (k+1) w_k \langle \Xi_{j-1} \Psi_j \Xi_{k+1} \Psi_k \rangle 
{\rm e}^{\frac{2\pi}{W}(2j-1)x} . \label{ie33}
\end{eqnarray}  
Considering the sum rule (\ref{sr32}) and following the preceding algebra 
leads to
\begin{eqnarray} 
\int_0^W {\rm d}y\, \cos\left(\frac{2\pi}{W}y\right) 
\left[ {\rm e}^{-\frac{2\pi}{W}d} U(x,d;y) - U(x,0;y) \right] 
+ \int_0^d {\rm d}x' \int_0^W {\rm d}y\, & & \nonumber \\ 
\times \cos\left(\frac{2\pi}{W}y\right)
{\rm e}^{-\frac{2\pi}{W}x'} S(x,x';y) \left\{ \frac{\partial}{\partial x'}
\left[ \beta v(x') \right] + \frac{\pi}{W} \left[ \Gamma(N-1) + 4 \right]
\right\} & & \nonumber \\ = - {\rm e}^{-\frac{2\pi}{W}x} 
\left[ \frac{\partial n(x)}{\partial x} - \frac{4\pi}{W} n(x) \right] .
\phantom{aa} \label{result2}
\end{eqnarray}  
For the one-body potential (\ref{vx}) it holds that
\begin{equation} \label{dodatok2}
\frac{\partial}{\partial x'} \left[ \beta v(x') \right] 
+ \frac{\pi}{W} \left[ \Gamma(N-1) + 4 \right] 
= 2\pi \Gamma \sigma + \frac{\pi}{W} (4-\Gamma) .
\end{equation}
The physical content of the exact sum rules (\ref{result1}), (\ref{dodatok1})
or (\ref{result2}), (\ref{dodatok2}) is not obvious for a finite value of $W$
due to the presence of the slowly changing factor $\cos(2\pi y/W)$ along 
the integration path over $y\in [0,W]$. 
On the other hand, these sum rules will be very useful in the limit
$W\to\infty$ to derive certain exact relations among relevant statistical 
quantities, see the next section.

\renewcommand{\theequation}{4.\arabic{equation}}
\setcounter{equation}{0}

\section{Asymptotic decay of pair correlations along the walls} \label{Sec:4}

\subsection{One-wall geometry}
Let us first consider the 2D geometry of one wall (infinite line) 
localized at $x=0$ and charged by the fixed charge density $\sigma e$.
The mobile counterions of charge $-e$ are constrained to the half-space $x>0$.
Their number density $n(x)$ fulfills the electroneutrality condition 
\begin{equation} \label{iee11}
\int_0^d {\rm d}x\, n(x) = \sigma . 
\end{equation}
 
Near a hard wall, the screening cloud around a test charge is asymmetric and
therefore the Ursell function exhibits a long-range (inverse-power law) decay 
at asymptotically large distances along the wall 
\cite{Jancovici82a,Jancovici82b,Usenko79}.
In 2D, the Ursell function between the points $(x,y)$ and $(x',y')$ behaves as
\begin{equation} \label{asymptotic}
U(x,x';\vert y-y'\vert) \mathop{\sim}_{\vert y-y'\vert\to\infty} 
\frac{f^{(1)}(x,x')}{(y-y')^2} ,
\end{equation} 
where the superscript $1$ in $f^{(1)}$ means that there is just one charged wall
at $x=0$.
The function $f^{(1)}(x,x')=f^{(1)}(x',x)$ obeys the sum rule 
\cite{Jancovici82b,Jancovici95,Jancovici01}
\begin{equation} \label{sumrule}
\int_0^{\infty} {\rm d}x \int_0^{\infty} {\rm d}x'\, f^{(1)}(x,x') =
- \frac{1}{2\pi^2\Gamma} .
\end{equation}
Note that this sum rule does not depend on $\sigma$.

Applying the M\"obius conformal transformation to particle coordinates in
a disc geometry and going from the disc to an infinite line 
\cite{Samaj15,Samaj16b}, it was found for the present model with counterions 
only that the function $f^{(1)}$ satisfies the following equations:
\begin{eqnarray}
f^{(1)}(x,0) & = & - \frac{1}{\pi} \left[ x \frac{\partial}{\partial x} n(x)
+ 2 n(x) \right] , \label{equation1} \\
f^{(1)}(x,0) & = & 2\pi\Gamma\sigma \int_0^{\infty} {\rm d}x'\, f^{(1)}(x,x') .
\label{equation2}
\end{eqnarray}
Applying $\int_0^{\infty}{\rm d}x$ to both sides of (\ref{equation2}) and
taking into account the sum rule (\ref{sumrule}), one finds that
\begin{equation} \label{equation3}
\int_0^{\infty} {\rm d}x'\, f^{(1)}(0,x') + \frac{\sigma}{\pi} = 0 .
\end{equation}
Finally, setting $x=0$ in (\ref{equation2}) the $f^{(1)}$-function with 
both points at the boundary is given by
\begin{equation} \label{equation4}
f^{(1)}(0,0) = - 2\Gamma \sigma^2 .
\end{equation}
 
For the single line charge density $\sigma e$, with counterions only, 
the particle density profile and the asymptotic function $f^{(1)}(x,x')$ 
were obtained in the PB limit $\Gamma\to 0$ \cite{Samaj13},
\begin{equation} \label{onewallgamma0}
n(x) = \frac{\sigma b}{(x+b)^2} , \qquad 
f^{(1)}(x,x') = - \frac{2}{\pi^2\Gamma} \frac{b^4}{(x+b)^3 (x'+b)^3} 
\end{equation}
with $b=1/(\Gamma\pi\sigma)$,
and at the free-fermion coupling $\Gamma=2$ \cite{Jancovici84,Samaj13},
\begin{equation} \label{onewallgamma2}
n(x) = \frac{1}{4\pi x^2} \left[ 1 - (1+4\pi\sigma x) {\rm e}^{-4\pi\sigma x}
\right] , \quad
f^{(1)}(x,x') = - 4 \sigma^2 {\rm e}^{-4\pi\sigma x} {\rm e}^{-4\pi\sigma x'} .
\end{equation} 
Note that while in the PB limit both $n(x)$ and $f^{(1)}(x,x')$ are long-ranged,
$n(x)$ is long-ranged but $f^{(1)}(x,x')$ is short-ranged at $\Gamma=2$. 
It is simple to check that the sum rules (\ref{equation1})-(\ref{equation4})
are fulfilled at both exactly solvable $\Gamma$'s.

\subsection{Two-walls geometry}
In the presence of two walls, the Ursell functions are supposed to
exhibit the same asymptotic behavior as in 
the one-wall case (\ref{asymptotic}), i.e.,
\begin{equation} \label{asymptotic2}
U(x,x';\vert y-y'\vert) \mathop{\sim}_{\vert y-y'\vert\to\infty} 
\frac{f^{(2)}(x,x')}{(y-y')^2} ,
\end{equation} 
where the superscript $2$ in $f^{(2)}$ means that there are two parallel 
charged walls, the one with the charge density $\sigma e$ at $x=0$ and 
the other with the charge density $\sigma' e$ at $x=d$.
The aim of this part is to investigate the thermodynamic $W\to\infty$ limit 
of the sum rules (\ref{result1}), (\ref{dodatok1}) and 
(\ref{result2}), (\ref{dodatok2}). 

Let us start with the analysis of the first sum rule (\ref{result1}), 
(\ref{dodatok1}) in the limit $W\to\infty$.
Using the zeroth-moment condition (\ref{zerothmoment}) and the WLMB equation 
(\ref{Wertheim}), the sum rule can be rewritten as
\begin{eqnarray} 
\int_0^W {\rm d}y\, \cos\left(\frac{2\pi}{W}y\right) 
\left( {\rm e}^{\frac{2\pi}{W}d} -1 \right) U(x,d;y) \nonumber \\
+ \int_0^W {\rm d}y\, \left[ \cos\left(\frac{2\pi}{W}y\right) - 1 \right] 
\left[ U(x,d;y) - U(x,0;y) \right]  \nonumber \\
- \left[ 2\pi \Gamma \sigma' + \frac{\pi}{W} (4-\Gamma) \right]
\int_0^d {\rm d}x' \int_0^W {\rm d}y\, 
\left[ \cos\left(\frac{2\pi}{W}y\right) {\rm e}^{\frac{2\pi}{W}x'} -1 \right]
S(x,x';y) \nonumber \\ 
= - \left( {\rm e}^{\frac{2\pi}{W}x} -1 \right) \frac{\partial n(x)}{\partial x} 
- \frac{4\pi}{W} n(x) {\rm e}^{\frac{2\pi}{W}x} . \nonumber \\ \label{result21}
\end{eqnarray}
In the limit $W\to\infty$, one expands 
\begin{equation} \label{exp1}
{\rm e}^{\frac{2\pi}{W}x} \sim 1 + \frac{2\pi}{W} x 
+ O\left( \frac{1}{W^2} \right) , \quad
\cos\left(\frac{2\pi}{W}y\right) \sim 1 
- \frac{1}{2!} \left( \frac{2\pi}{W} \right)^2 y^2 +
O\left( \frac{1}{W^4} \right) .  
\end{equation}
The integrals of the Ursell functions $U$ (or the structure function $S$)
over $y$ can be done in the following way
\begin{eqnarray}
\int_0^W {\rm d}y\, \left[ \cos\left(\frac{2\pi}{W}y\right) - 1 \right] 
U(x,x';y) & \displaystyle{\mathop{\sim}_{W\to\infty}} & - \frac{1}{2!} 
\left( \frac{2\pi}{W} \right)^2 \int_0^W {\rm d}y\, y^2
\frac{f^{(2)}(x,x')}{y^2} \nonumber \\
& = & - \frac{2\pi^2}{W} f^{(2)}(x,x') . \label{exp2}
\end{eqnarray} 
Comparing in (\ref{result21}) the terms proportional to $1/W$ 
implies the equality among the 2D statistical quantities:
\begin{eqnarray}
d \int_{-\infty}^{\infty} {\rm d}y\, U(x,d;y) - 2\pi\Gamma\sigma' 
\int_0^d {\rm d}x'\, x' \int_{-\infty}^{\infty} {\rm d}y\, S(x,x';y)
\nonumber \\ + \pi \left[ f^{(2)}(x,0) - f^{(2)}(x,d) \right] 
+ 2\pi^2 \Gamma \sigma' \int_0^d {\rm d}x'\, f^{(2)}(x,x') \nonumber \\
= - \left[ x \frac{\partial n(x)}{\partial x} + 2 n(x) \right] .
\label{vys1}
\end{eqnarray}

We proceed analogously with the second sum rule (\ref{result2}), 
(\ref{dodatok2}).
With the aid of the zeroth-moment condition (\ref{zerothmoment}) and 
the WLMB equation (\ref{Wertheim}), one gets
\begin{eqnarray} 
\int_0^W {\rm d}y\, \cos\left(\frac{2\pi}{W}y\right) 
\left( {\rm e}^{-\frac{2\pi}{W}d} -1 \right) U(x,d;y) \nonumber \\
+ \int_0^W {\rm d}y\, \left[ \cos\left(\frac{2\pi}{W}y\right) - 1 \right] 
\left[ U(x,d;y) - U(x,0;y) \right]  \nonumber \\
+ \left[ 2\pi \Gamma \sigma + \frac{\pi}{W} (4-\Gamma) \right]
\int_0^d {\rm d}x' \int_0^W {\rm d}y\, 
\left[ \cos\left(\frac{2\pi}{W}y\right) {\rm e}^{-\frac{2\pi}{W}x'} -1 \right]
S(x,x';y) \nonumber \\ 
= - \left( {\rm e}^{-\frac{2\pi}{W}x} -1 \right) \frac{\partial n(x)}{\partial x} 
+ \frac{4\pi}{W} n(x) {\rm e}^{-\frac{2\pi}{W}x} . \nonumber \\ \label{result22}
\end{eqnarray}
Using (\ref{exp1}) and (\ref{exp2}) and comparing in (\ref{result22}) 
the terms proportional to $1/W$ leads to the equality 
\begin{eqnarray}
- d \int_{-\infty}^{\infty} {\rm d}y\, U(x,d;y) - 2\pi\Gamma\sigma 
\int_0^d {\rm d}x'\, x' \int_{-\infty}^{\infty} {\rm d}y\, S(x,x';y)
\nonumber \\ + \pi \left[ f^{(2)}(x,0) - f^{(2)}(x,d) \right] 
- 2\pi^2 \Gamma \sigma \int_0^d {\rm d}x'\, f^{(2)}(x,x') \nonumber \\
= x \frac{\partial n(x)}{\partial x} + 2 n(x) . \label{vys2}
\end{eqnarray}

The crucial 2D Eqs. (\ref{vys1}) and (\ref{vys2}) are valid for any coupling
$\Gamma = 2\gamma$ with $\gamma$ a positive integer.
It is natural to extend their validity to all real $\Gamma$ in the fluid region.
We can obtain a couple of simpler relations by considering specific 
combinations of the two equations.
The summation of Eqs. (\ref{vys1}) and (\ref{vys2}) results in
\begin{eqnarray}
- \Gamma(\sigma+\sigma') \int_0^d {\rm d}x'\, x' 
\int_{-\infty}^{\infty} {\rm d}y\, S(x,x';y) + f^{(2)}(x,0) - f^{(2)}(x,d) 
\nonumber \\ 
+ \pi \Gamma (\sigma'-\sigma)  \int_0^d {\rm d}x'\, f^{(2)}(x,x') 
= 0 . \label{rovnica1}
\end{eqnarray}
The subtraction of Eqs. (\ref{vys1}) and (\ref{vys2}) implies that
\begin{eqnarray}
\frac{d}{\pi} \int_{-\infty}^{\infty} {\rm d}y\, U(x,d;y) 
+ \Gamma(\sigma-\sigma') \int_0^d {\rm d}x'\, x' 
\int_{-\infty}^{\infty} {\rm d}y\, S(x,x';y) \nonumber \\ 
+ \pi \Gamma (\sigma+\sigma') \int_0^d {\rm d}x'\, f^{(2)}(x,x') 
= -\frac{1}{\pi} \left[ x \frac{\partial n(x)}{\partial x} 
+ 2 n(x) \right] . \label{rovnica2}
\end{eqnarray} 

Integrating both sides of Eq. (\ref{rovnica2}) over $x\in [0,d]$, 
the integration of $S(x,x';y)$ over $x'$ can be interchanged with the
integration over $x$ for a finite value of $d$ and the corresponding term 
vanishes due to the counterpart of the zeroth-moment condition 
(\ref{zerothmoment})
\begin{equation} \label{zerothmoment1}
\int_0^d {\rm d}x' \int_{-\infty}^{\infty} {\rm d}y\, S(x,x';y) = 0 . 
\end{equation}
We emphasize that the interchange of the integrations cannot be performed
in the one-wall limit $d\to\infty$ because, as is known, the integral over 
$x'$ is not absolutely convergent.
The integral over $x$ of the rhs of Eq. (\ref{rovnica2}) can be simplified
by applying the integration by parts:
\begin{equation}
- \frac{1}{\pi} \int_0^d {\rm d}x\, 
\left[ x \frac{\partial n(x)}{\partial x} + 2 n(x) \right] 
= -\frac{d}{\pi} n(d) - \frac{1}{\pi} \int_0^d {\rm d}x\, n(x) .
\end{equation}  
The term $(d/\pi) n(d)$ can be paired with the one
$(d/\pi) \int_0^d {\rm d}x \int_{-\infty}^{\infty} {\rm d}y\, U(x,d;y)$
to get 0 due to (\ref{zerothmoment1}).
Expressing the integral $\int_0^d {\rm d}x\, n(x)$ by using (\ref{iee11}),
we end up with the sum rule
\begin{equation} \label{sumrule2}
\int_0^d {\rm d}x \int_0^d {\rm d}x'\, f^{(2)}(x,x') = - \frac{1}{\pi^2\Gamma} .
\end{equation}
Note that this sum rule does not depend neither on the surface charge
densities $\sigma e$ and $\sigma' e$, nor on the distance between the walls
$d$.  
In comparison with the analogous formula for the one-wall geometry
(\ref{sumrule}), the factor $1/2$ is missing on the rhs of (\ref{sumrule2}).
To explain this fact, let us consider the special limit $d\to\infty$ of two
independent walls when, for finite values of $x$ and $x'$,
\begin{equation}
f^{(2)}(x,x') \mathop{\sim}_{d\to\infty} 
f^{(1)}(x,x';\sigma) + f^{(1)}(d-x,d-x';\sigma') . 
\end{equation}
Integrating over coordinates $x$ and $x'$ and changing the integration 
variables to $d-x, d-x'$ when integrating the second term, one finds that 
the double integral of $f^{(2)}$ must be twice the double integral 
of $f^{(1)}$; note that the argument works because the sum rule (\ref{sumrule2})
does not depend on $d$. 

Another sum rule can be obtained by integrating both sides of 
Eq. (\ref{rovnica1}) over $x\in [0,d]$. 
The integral of $S(x,x';y)$ vanishes once more and using (\ref{sumrule2})
one gets
\begin{equation} \label{sepa}
\int_0^d {\rm d}x'\, f^{(2)}(0,x') + \frac{\sigma}{\pi} = 
\int_0^d {\rm d}x'\, f^{(2)}(d,x') + \frac{\sigma'}{\pi} . 
\end{equation}
In other words, for each of the walls the combination (\ref{equation3}), 
which is equal to zero for one-wall geometry, acquires the same value in 
the two-walls geometry.

\subsection{Small-distance behavior}
In the limit $d\to 0$, the particle density, the pressure and the
asymptotic function $f^{(2)}(x,x')$ exhibit singularities.
As is evident from the electroneutrality condition (\ref{iee11}), the
particle density behaves as
\begin{equation}
n(x) \mathop{\sim}_{d\to 0} \frac{\sigma+\sigma'}{d} .
\end{equation} 
Since the pressure is determined by the contact particle density, we have
likewisely
\begin{equation} \label{Pasymptotic}
\beta P \mathop{\sim}_{d\to 0} \frac{\sigma+\sigma'}{d} .
\end{equation} 
The asymptotic function $f^{(2)}(x,x')$ is searched in the ansatz form
\begin{equation}
f^{(2)}(x,x') \mathop{\sim}_{d\to 0} \frac{1}{d^2} \left[ a + b(x+x') 
+ \cdots \right] .
\end{equation}
Inserting this expansion into the sum rules (\ref{sumrule2}) and (\ref{sepa}),
the expansion coefficients are found to be
\begin{equation}
a = - \frac{1}{\pi^2 \Gamma} , \qquad b = \frac{1}{\pi} (\sigma-\sigma') .
\end{equation}

\renewcommand{\theequation}{5.\arabic{equation}}
\setcounter{equation}{0}

\section{The free-fermion coupling} \label{Sec:5}
At the free-fermion coupling $\Gamma=2$ $(\gamma=1)$, the composite operators
$\Xi_j$ and $\Psi_j$ become the ordinary anticommuting variables 
$\xi_j$ and $\psi_j$, respectively.
Having the diagonalized action $S=\sum_{j=0}^{N-1} \xi_j w_j \psi_j$, 
the integral over anticommuting variables (\ref{antipart}) reads as
\begin{equation}
Q_N(1) = \prod_{j=0}^{N-1} w_j ,
\end{equation}
where the interaction strengths (\ref{wj}) take for $\gamma=1$ the form
\begin{equation} \label{wjprime}
w_j = \frac{1-\exp\left[ - \frac{4\pi d}{W}\left(j-\sigma' W
+\frac{1}{2}\right)\right]}{j-\sigma' W+\frac{1}{2}} .
\end{equation}
The simplest correlators of anticommuting variables are given by
\begin{equation} \label{onecorrelator}
\langle \xi_j\psi_j \rangle = \frac{1}{w_j} , \qquad j=0,1,\ldots,N-1.
\end{equation}
More complicated correlators can be obtained by using the Wick theorem.
Like for instance,
\begin{equation} \label{twocorrelator}
\langle \xi_j\psi_k \xi_{j'} \psi_{k'} \rangle 
= \frac{1}{w_jw_{j'}} \left( \delta_{jk}\delta_{j'k'} -
\delta_{jk'}\delta_{j'k} \right) .
\end{equation}

\subsection{Particle density and pressure}
Inserting (\ref{onecorrelator}) into the formula (\ref{antione}) for the
particle density, one gets
\begin{eqnarray} 
n(x) & =  & \frac{4\pi}{W^2} \sum_{j=0}^{N-1} 
\frac{j-\sigma' W+\frac{1}{2}}{1-\exp\left[ - \frac{4\pi d}{W}
\left(j-\sigma' W+\frac{1}{2}\right)\right]} \nonumber \\ & & \times
\exp\left[ - \frac{4\pi x}{W} \left( j -\sigma' W 
+ \frac{1}{2} \right) \right] . \label{antioneone}
\end{eqnarray}

To obtain the explicit results for the 2D geometry of two parallel lines 
charged by the line charge densities $\sigma e$ and $\sigma' e$, 
at distance $d$ with counterions only in between, 
we consider the thermodynamic limit $N,W\to\infty$, keeping the ratio 
$N/W = \sigma+\sigma'$ fixed.
Choosing $t=\left( j-W\sigma'+\frac{1}{2}\right)/N$ as the continuous variable, 
the particle density (\ref{antioneone}) can be expressed as
\begin{eqnarray}
n(x) & = & 4\pi \left( \frac{N}{W} \right)^2 
\int_{-\frac{\sigma'}{\sigma+\sigma'}}^{\frac{\sigma}{\sigma+\sigma'}} {\rm d}t\, 
t \frac{{\rm e}^{-4\pi(\sigma+\sigma')t x}}{1-{\rm e}^{-4\pi(\sigma+\sigma')t d}}
\nonumber \\ & = & \frac{1}{4\pi} \int_{-4\pi\sigma'}^{4\pi\sigma} {\rm d}s\, s
\frac{{\rm e}^{-s x}}{1-{\rm e}^{-s d}} \nonumber \\ & 
= & n_0(x;\sigma) + n_0(d-x;\sigma') , \label{separate}
\end{eqnarray}
where
\begin{equation} \label{separate1}
n_0(x;\sigma) = \frac{1}{4\pi} \int_0^{4\pi\sigma} {\rm d}s\, s
\frac{{\rm e}^{-s x}}{1-{\rm e}^{-s d}} 
\end{equation}
is the density of counterions between two parallel lines, the one at $x=0$ 
charged with the line charge density $\sigma e$ and the neutral one at $x=d$.
It is trivial to verify that $\int_0^d {\rm d}x\, n_0(x;\sigma) = \sigma$ 
as it should be.
The separation form of the density profile (\ref{separate}) as the sum of two 
terms, the one depending only on $\sigma$ and the other depending only on
$\sigma'$, is the special feature of the free-fermion point.

The pressure is given by the contact relations (\ref{PN}).
Choosing the one $\beta P_N = n(0) - 2\pi \sigma^2$, from (\ref{antioneone}) 
one gets
\begin{equation} 
\beta P_N(d) =  \frac{4\pi}{W^2} \sum_{j=0}^{N-1} 
\frac{j-\sigma' W+\frac{1}{2}}{1-\exp\left[ - \frac{4\pi d}{W}
\left(j-\sigma' W+\frac{1}{2}\right)\right]} - 2\pi \sigma^2 .
\end{equation}
The pressure is expected to vanish in the limit of an infinite distance 
between the lines, but this is not the case for a finite odd value of $N$.
Let us document this fact on the pair of symmetrically charged lines 
$\sigma'=\sigma$, i.e. $N=2\sigma W$.
In the limit $d\to\infty$, only terms with $j-\sigma W +\frac{1}{2}>0$
contribute to the pressure:
\begin{equation} 
\lim_{d\to\infty} \beta P_N(d) =  \frac{4\pi}{W^2} 
\sum_{j=0\atop j>W\sigma-\frac{1}{2}}^{N-1} 
\left( j - \sigma W +\frac{1}{2} \right) - 2\pi \sigma^2 .
\end{equation} 
If $N$ is an even integer, $\sigma W$ is an integer and one has
\begin{equation} 
\lim_{d\to\infty} \beta P_N(d) =  \frac{4\pi}{W^2} 
\sum_{j=\sigma W}^{2\sigma W-1} 
\left( j - \sigma W +\frac{1}{2} \right) - 2\pi \sigma^2 = 0.
\end{equation} 
If $N$ is an odd integer, $\sigma W$ is a half-integer and one has
\begin{equation} 
\lim_{d\to\infty} \beta P_N(d) =  \frac{4\pi}{W^2} 
\sum_{j=\sigma W+\frac{1}{2}}^{2\sigma W-1} 
\left( j - \sigma W +\frac{1}{2} \right) - 2\pi \sigma^2 = 
- \frac{2\pi\sigma^2}{N^2} .
\end{equation}  
A nonzero asymptotic force for an odd number of counterions between two 
charges occurs also in 1D \cite{Tellez15,Varela17}.

Let us express the pressure by using a symmetric combination of 
the contact relations (\ref{PN}),
\begin{equation}
\beta P(d) = \frac{1}{2} \left[ n(0) + n(d) \right] 
- \pi \left( \sigma^2 + \sigma'^2 \right) .
\end{equation}
In the thermodynamic limit (pure 2D geometry), using the explicit results 
(\ref{separate}) and (\ref{separate1}) one obtains the separable solution
\begin{equation} \label{pressure}
\beta P(d) = \beta P(d;\sigma) + \beta P(d;\sigma') ,
\end{equation}
where
\begin{equation} \label{separatepressure}
\beta P(d;\sigma) = \frac{1}{2\pi d^2} \int_0^{2\pi\sigma d} {\rm d}t\,
\frac{t}{\sinh t} {\rm e}^{-t} 
\end{equation}
is the pressure between two parallel lines, the one at $x=0$ charged with 
the line charge density $\sigma e$ and another neutral one at $x=d$.

For the studied case of like-charged lines ($0<\sigma'<\sigma$), 
$\beta P$ is always positive, i.e. the two lines repel each other 
for an arbitrary distance $d$.
Using the substitution $t=d s$ in (\ref{separatepressure}) it can be shown
that 
\begin{equation}
\frac{\partial}{\partial d} \beta P = - \frac{1}{2\pi} 
\left( \int_0^{2\pi\sigma} + \int_0^{2\pi\sigma'} \right) {\rm d}s
\left[ \frac{s}{\sinh(d s)} \right]^2 < 0 .
\end{equation}
The pressure (\ref{pressure}) diverges at small distances in agreement
with the general formula (\ref{Pasymptotic}) and decays monotonously to 
0 at $d\to\infty$.
If $0<\sigma'\le \sigma$, the asymptotic decay 
\begin{equation}
\beta P(d) \mathop{\sim}_{d\to\infty} \frac{1}{\pi d^2} 
\int_0^{\infty} {\rm d}t\, \frac{t}{\sinh t} {\rm e}^{-t} 
= \frac{\pi}{12} \frac{1}{d^2} 
\end{equation}
is universal in the sense that the prefactor to $1/d^2$ is independent of 
the (positive) line charge densities.

\begin{figure}[tbh]
\begin{center}
\includegraphics[width=0.8\textwidth,clip]{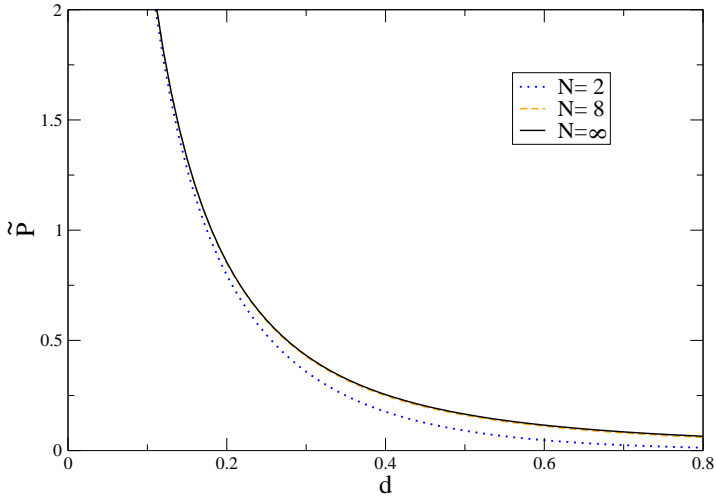}
\caption{The dimensionless pressure $\tilde{P}$ versus the dimensionless
distance $d$ (measured in units of $\sigma^{-1}$) between the symmetrically 
charged walls with the asymmetry parameter $\eta=1$ for 
the free-fermion coupling $\Gamma=2$ $(\gamma=1)$.
The dotted (blue) curve corresponds to $N=2$ particles, the dashed (orange) 
curve to $N=8$ and the solid (black) curve to the thermodynamic limit
$N\to\infty$.}
\label{Fig:2}
\end{center}
\end{figure}

The data for the dimensionless pressure $\tilde{P}$ (\ref{Ptilde}) 
versus the dimensionless distance $d$ (measured in units of $\sigma^{-1}$) 
between the symmetrically charged walls ($\eta=1$), calculated by using
Eqs. (\ref{pressure}) and (\ref{separatepressure}), are pictured 
in Fig. \ref{Fig:2}.
The dotted curve corresponds to $N=2$ particles, the dashed curve to $N=8$ 
particles and the solid curve to $N\to\infty$ particles.
The $N=8$ and $N\to\infty$ curves are almost indistinguishable, 
so that the equation of state for $N=8$ particles is practically identical 
to the one in the thermodynamic limit.
The same behavior is observed for all values of the asymmetry parameter
$\eta\in [0,1]$. 
We anticipate that the quick convergence of data with the particle number
$N$, observed for $\Gamma=2$, is maintained also for the higher couplings
$\Gamma$. 

\subsection{Two-body density}
Inserting the correlators (\ref{twocorrelator}) into the formula 
for the two-body density (\ref{antitwo}), the Ursell function (\ref{Ursell})
is expressible as
\begin{eqnarray}
U(z_1,\bar{z}_1;z_2,\bar{z}_2) & = & - w_{\rm ren}(x_1) w_{\rm ren}(x_2)
\sum_{j,j'=0}^{N-1} \frac{1}{w_j w_{j'}} \nonumber \\ & & \times
\exp\left[ -\frac{2\pi}{W} j\left(z_1+\bar{z}_2\right) \right] 
\exp\left[ -\frac{2\pi}{W} j'\left(\bar{z}_1+z_2\right) \right] .
\phantom{aaa} \label{Ursell2} 
\end{eqnarray}  
Let us denote $y\equiv y_1-y_2$ and note that our $w_{\rm ren}(x)$ 
(\ref{onebody}) with $\beta v(x)$ given by (\ref{vx}) satisfies
the relation
\begin{equation}
w_{\rm ren}(x_1) w_{\rm ren}(x_2) = 
w_{\rm ren}\left(\frac{x_1+x_2}{2}+{\rm i}\frac{y}{2} \right) 
w_{\rm ren}\left(\frac{x_1+x_2}{2}-{\rm i}\frac{y}{2} \right) . 
\end{equation} 
The Ursell function (\ref{Ursell2}) is thus expressible in terms
of the particle density as
\begin{equation} \label{Ursell3}
U(x_1,x_2;y) = - n\left(\frac{x_1+x_2}{2}+{\rm i}\frac{y}{2} \right) 
n\left(\frac{x_1+x_2}{2}-{\rm i}\frac{y}{2} \right) . 
\end{equation}

The availability of the explicit formula for the Ursell function
(\ref{Ursell3}) enables us to investigate the effect of the two-wall 
geometry on the prefactor function $f^{(2)}(x_1,x_2)$ defined by
Eq. (\ref{asymptotic2}).
The particle density (\ref{separate}) consists of two similar terms.
The first term (\ref{separate1}) can be expanded as follows 
\begin{eqnarray} 
n_0(x;\sigma) & = & -\frac{1}{4\pi} \frac{\partial}{\partial x}
\int_0^{4\pi\sigma} {\rm d}s\, {\rm e}^{-s x} \sum_{j=0}^{\infty} {\rm e}^{-j s d} 
\nonumber \\ & = & -\frac{1}{4\pi} \frac{\partial}{\partial x}
\sum_{j=0}^{\infty} \frac{1-{\rm e}^{-4\pi\sigma(x+jd)}}{x+jd} \nonumber \\ 
& = & \sum_{j=0}^{\infty} \left[ \frac{1}{4\pi} 
\frac{1-{\rm e}^{-4\pi\sigma(x+jd)}}{(x+jd)^2} - \sigma
\frac{{\rm e}^{-4\pi\sigma(x+jd)}}{x+jd} \right] . \label{separate2}
\end{eqnarray}
Substituting $x$ by $(x_1+x_2\pm{\rm i}y)/2$ in the last line of this
expression and considering the limit $y\to\infty$, one finds that
\begin{equation}
n_0\left(\frac{x_1+x_2}{2}\pm{\rm i}\frac{y}{2};\sigma\right)
\mathop{\sim}_{y\to\infty} \pm {\rm i} \frac{2\sigma}{y}
\frac{{\rm e}^{-2\pi\sigma[(x_1+x_2)\pm{\rm i}y]}}{1-{\rm e}^{-4\pi\sigma d}} .
\end{equation}
Analogously, 
\begin{equation}
n_0\left(d-\frac{x_1+x_2}{2}\mp{\rm i}\frac{y}{2};\sigma'\right)
\mathop{\sim}_{y\to\infty} \mp {\rm i} \frac{2\sigma'}{y}
\frac{{\rm e}^{-2\pi\sigma'[2d-(x_1+x_2)\mp{\rm i}y]}}{1-{\rm e}^{-4\pi\sigma' d}} .
\end{equation}
We conclude that
\begin{eqnarray}
n\left(\frac{x_1+x_2}{2}\pm{\rm i}\frac{y}{2} \right) 
& \displaystyle{\mathop{\sim}_{y\to\infty}} & \pm {\rm i} \frac{2}{y} 
\left\{ \sigma 
\frac{{\rm e}^{-2\pi\sigma[(x_1+x_2)\pm{\rm i}y]}}{1-{\rm e}^{-4\pi\sigma d}} 
\right. \nonumber \\ & & \left. \qquad - \sigma' 
\frac{{\rm e}^{-2\pi\sigma'[2d-(x_1+x_2)\mp{\rm i}y]}}{1-{\rm e}^{-4\pi\sigma' d}} 
\right\} .
\end{eqnarray}
The Ursell function (\ref{Ursell3}) then exhibits the asymptotic behavior
\begin{eqnarray}
U(x_1,x_2) & \displaystyle{\mathop{\sim}_{y\to\infty}} & 
- \frac{4}{y^2} \Bigg\{ \frac{\sigma^2 {\rm e}^{-4\pi\sigma(x_1+x_2)}}{
\left(1-{\rm e}^{-4\pi\sigma d}\right)^2} + \frac{\sigma'^2 
{\rm e}^{-4\pi\sigma'[2d-(x_1+x_2)]}}{ \left(1-{\rm e}^{-4\pi\sigma' d}\right)^2}
\nonumber \\ & &  
- \frac{2\sigma\sigma' {\rm e}^{-2\pi\sigma(x_1+x_2)-2\pi\sigma'[2d-(x_1+x_2)]}}{
\left(1-{\rm e}^{-4\pi\sigma d}\right)\left(1-{\rm e}^{-4\pi\sigma' d}\right)} 
\cos\left[2\pi(\sigma+\sigma')y\right] \Bigg\} . \nonumber \\
& & \label{twowall}
\end{eqnarray}
This asymptotic result is of type (\ref{asymptotic2}) with
\begin{equation} \label{twowallf}
f^{(2)}(x_1,x_2) = - 4 \left\{ \frac{\sigma^2 {\rm e}^{-4\pi\sigma(x_1+x_2)}}{
\left(1-{\rm e}^{-4\pi\sigma d}\right)^2} + \frac{\sigma'^2 
{\rm e}^{-4\pi\sigma'[2d-(x_1+x_2)]}}{ \left(1-{\rm e}^{-4\pi\sigma' d}\right)^2}
\right\} ;
\end{equation}
note that the oscillating term in (\ref{twowall}) does not contribute
to this function.
It is simple to check that $f^{(2)}(x,x')$ satisfies both sum rules 
(\ref{sumrule2}) and (\ref{sepa}).  
In the limit $d\to\infty$, keeping the coordinates $x$ and $x'$ finite,
the two-wall formula (\ref{twowallf}) reduces itself to the semi-infinite 
one-wall result (\ref{onewallgamma2}) as it should be.
For a finite $d$, $f^{(2)}(x,x')$ cannot be written in the factorized form 
$- g(x_1) g(x_2)$ as in the one-wall case.

Using the formulas
\begin{eqnarray}
\int_{-\infty}^{\infty} {\rm d}y\, U(x,x';y) & = & - \frac{1}{4\pi}
\int_{-4\pi\sigma'}^{4\pi\sigma} {\rm d}s\, s^2 
\frac{{\rm e}^{-s(x+x')}}{\left( 1-{\rm e}^{-s d}\right)^2} , \\
\int_0^d {\rm d}x'\, x' \int_{-\infty}^{\infty} {\rm d}y\, S(x,x';y) 
& = & - \frac{\sigma {\rm e}^{-4\pi\sigma x}}{1-{\rm e}^{-4\pi\sigma d}} +
\frac{\sigma' {\rm e}^{-4\pi\sigma'(d-x)}}{1-{\rm e}^{-4\pi\sigma' d}} ,
\end{eqnarray}
with the explicit forms of the particle density $n(x)$ (\ref{separate})
and the function $f^{(2)}(x,x')$ (\ref{twowallf}), it can be straightforwardly 
shown that Eqs. (\ref{rovnica1}) and (\ref{rovnica2}) hold.

\renewcommand{\theequation}{6.\arabic{equation}}
\setcounter{equation}{0}

\section{Couplings $\Gamma=2\gamma$ $(\gamma=2,3,\ldots)$} \label{Sec:6}
The two-wall problem can be solved also for higher couplings
$\Gamma=2\gamma$ $(\gamma=2,3,\ldots)$ by expressing $Q_N(\gamma)$, 
the integral over anticommuting variables (\ref{antipart}), as a function of 
the interaction strengths $w_j$ $[j=0,1,\ldots,\gamma(N-1)]$.
This can be done for lower values of $N$ \cite{Samaj04a}.
For $\gamma=2$, one has
\begin{eqnarray}
Q_2(2) & = & w_0 w_2 + 2 w_1^2 , \nonumber \\
Q_3(2) & = & w_0 w_2 w_4 + 2 w_0 w_3^2 + 2 w_1^2 w_4 + 4 w_1 w_2 w_3 + 6 w_2^3 , 
\nonumber \\
Q_4(2) & = & w_0 w_2 w_4 w_6 + 2 w_0 w_2 w_5^2 + 2 w_0 w_3^2 w_6 
+ 2 w_1^2 w_4 w_6 \nonumber \\ & &
+ 4 w_0 w_3 w_4 w_5 + 4 w_1 w_2 w_3 w_6 + 4 w_1 w_2 w_4 w_5 + 4 w_1^2 w_5^2 
\nonumber \\ & & + 4 w_2^2 w_4^2 + 6 w_0 w_4^3 + 6 w_2^3 w_6 + 8 w_1 w_3 w_4^2 
\nonumber \\ & & + 8 w_1 w_3^2 w_5 + 8 w_2^2 w_3 w_5 
+ 18w_2 w_3^2 w_4 + 24 w_3^4 , \label{gamma2}
\end{eqnarray}
etc. 
For $\gamma=3$, one has
\begin{eqnarray}
Q_2(3) & = & w_0 w_3 + 3^2 w_1 w_2 , \nonumber \\
Q_3(3) & = & w_0 w_3 w_6 + 3^2 w_0 w_4 w_5 + 3^2 w_1 w_2 w_6 \nonumber \\ & &
+ 6^2 w_1 w_3 w_5 + 15^2 w_2 w_3 w_4 , \nonumber \\
Q_4(3) & = & w_0 w_3 w_6 w_9 + 3^2 w_0 w_3 w_7 w_8 + 3^2 w_0 w_4 w_5 w_9 
+ 3^2 w_1 w_2 w_6 w_9 \nonumber \\ & &
+ 6^2 w_0 w_4 w_6 w_8 + 6^2 w_1 w_3 w_5 w_9  + 6^2 w_2 w_3 w_6 w_7 
+ 9^2 w_1 w_2 w_7 w_8 \nonumber \\ & &
+ 9^2 w_1 w_4 w_5 w_8 + 12^2 w_1 w_3 w_6 w_8 + 15^2 w_0 w_5 w_6 w_7 + 
15^2 w_2 w_3 w_4 w_9 \nonumber \\ & & 
+ 27^2 w_1 w_4 w_6 w_7 + 27^2 w_2 w_3 w_5 w_8 
+ 45^2 w_2 w_4 w_5 w_7 + 105^2 w_3 w_4 w_5 w_6 , \nonumber \\ & &  
\label{gamma3} 
\end{eqnarray}
etc.
\footnote{The explicit formulas for $Q_N(2)$ and $Q_N(3)$ up to $N=10$ will be
sent upon request by the author.} 
For the specific case of the one-component system constrained 
to a unit circle with all $w_j=1$ it was proved that \cite{Mehta}
\begin{equation}
Q_N(\gamma) = \frac{(\gamma N)!}{(\gamma!)^N N!} \qquad
\mbox{for all $w_j=1$.}
\end{equation}
The expressions (\ref{gamma2}) and (\ref{gamma3}) pass this test of
validity.

\begin{figure}[tbh]
\begin{center}
\includegraphics[width=0.8\textwidth,clip]{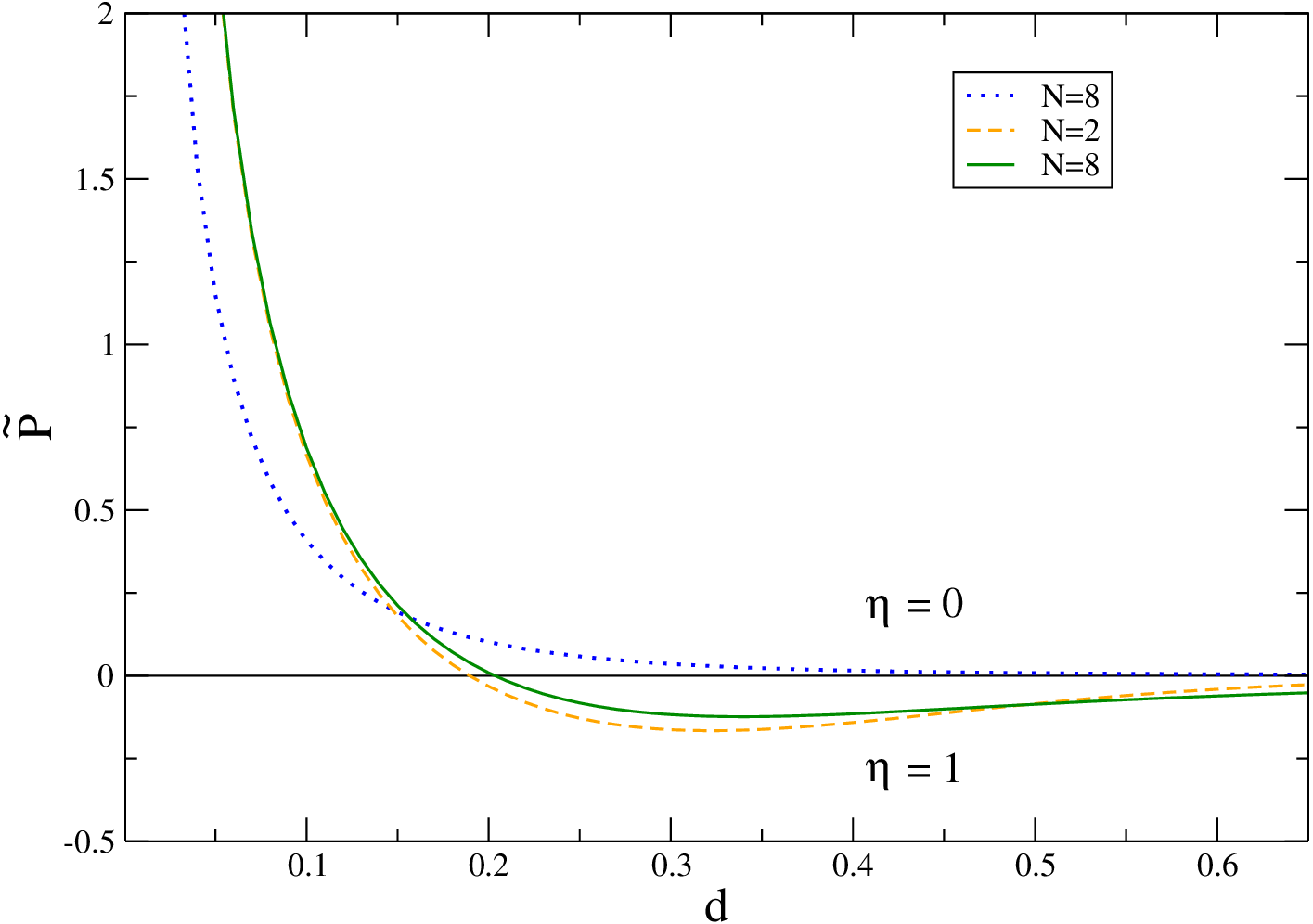}
\caption{The pressure $\tilde{P}$ versus the distance $d$ 
for the coupling $\Gamma=4$ $(\gamma=2)$.
The dotted (blue) curve corresponds to the asymmetry parameter $\eta=0$
and $N=8$ particles, the dashed (orange) curve to $\eta=1$ and $N=2$,
the solid (green) curve to $\eta=1$ and $N=8$.}
\label{Fig:3}
\end{center}
\end{figure}

Another possibility is to express explicitly for $N=2$ particles
$Q_2(\gamma)$ with an arbitrary integer value of $\gamma$:
\begin{equation} \label{Q2}
Q_2(\gamma) = \frac{1}{2} \sum_{j=0}^{\gamma} {\gamma\choose j}^2
w_j w_{\gamma-j} .  
\end{equation}

The free energy $F_N(\gamma)$ is expressed in terms of $Q_N(\gamma)$
in Eq. (\ref{free}), the pressure is calculated by using Eq. (\ref{PNN})
and the interaction strengths are given by (\ref{wj}).
Even for a relatively large number of particles $N=8$, the calculation of
the pressure from the exact formulas by using {\it Mathematica} takes 
a few seconds of CPU time on the standard PC.
 
For the coupling $\Gamma=4$, the exact data for the (dimensionless) 
pressure $\tilde{P}$ as the function of the (dimensionless) distance 
between the walls $d$ are presented in Fig. \ref{Fig:3}.
If the wall at $x=d$ does not carry any charge, i.e. $\sigma'=0$ or $\eta=0$,
$\tilde{P}$ is always positive for finite $d$, in agreement with 
the remark after Eq. (\ref{PN}), and its decay to zero 
at asymptotically large $d$ is monotonous for any number of particles $N$; 
this fact is documented for $N=8$ in Fig. \ref{Fig:3} by the dotted curve.
On the other hand, in the symmetric case $\eta=1$, for any $N$ 
there is a point at which $\tilde{P}$ intersects the $d$-axis 
and the pressure becomes negative, reaches a global minimum and
stays to be negative up to $d\to\infty$. 
For the particle numbers $N=2$ and $N=8$ this fact is documented in 
Fig. \ref{Fig:3} by the dashed and solid curves, respectively; 
note that the two curves are very close to one another which confirms 
the expected quick convergence of data with increasing $N$.
We conclude that the attraction phenomenon arises in 2D starting from 
a relatively small coupling constant $\Gamma$, somewhere between 2 and 4.

\begin{figure}[tbh]
\begin{center}
\includegraphics[width=0.8\textwidth,clip]{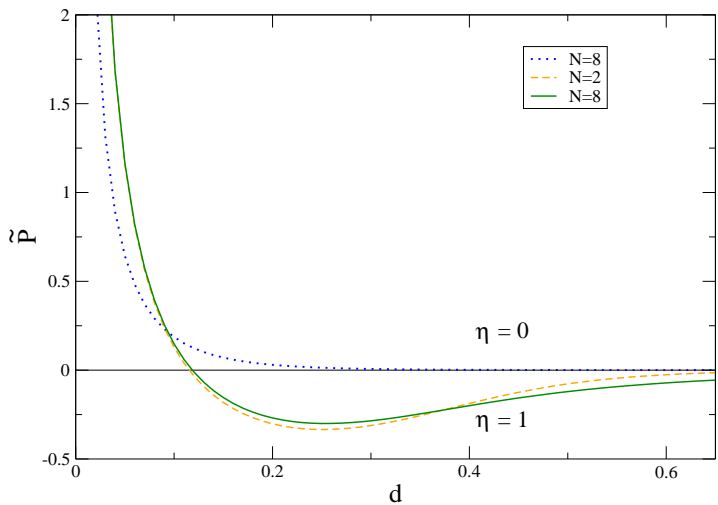}
\caption{The pressure $\tilde{P}$ versus the distance between the walls $d$ 
for the coupling $\Gamma=6$ $(\gamma=3)$.
The dotted (blue) curve corresponds to the asymmetry parameter $\eta=0$
and $N=8$ particles, the dashed (orange) curve to $\eta=1$ and $N=2$,
the solid (green) curve to $\eta=1$ and $N=8$.}
\label{Fig:4}
\end{center}
\end{figure}

The exact data for the pressure $\tilde{P}$ versus the distance $d$ 
for the coupling $\Gamma=6$ are presented in Fig. \ref{Fig:4}.
As before, the dotted curve corresponds to $\eta=0$ and $N=8$, 
the dashed curve to $\eta=1$ and $N=2$ and the solid curve to $\eta=1$ 
and $N=8$.
The results for $\Gamma=4$ and $\Gamma=6$ are similar qualitatively,
the global minima for $\eta=1$ are quantitatively more profound at $\Gamma=6$. 

\begin{figure}[tbh]
\begin{center}
\includegraphics[width=0.8\textwidth,clip]{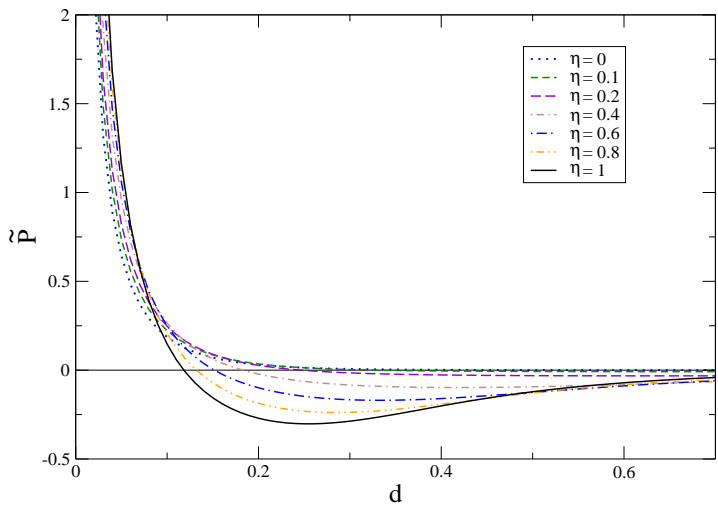}
\caption{The pressure $\tilde{P}$ versus the distance between the walls $d$ 
for the coupling $\Gamma=6$ and $N=6$ particles.
The asymmetry parameter $\eta$ takes successively the values 
$0,0.1,0.2,0.4,0.6,0.8,1$, the corresponding curves go at $d=0.3$ from up 
to down.}
\label{Fig:5}
\end{center}
\end{figure}

A natural question is at which value of the asymmetry parameter $\eta$
the monotonous decay of the positive $\tilde{P}$, observed at $\eta=0$, 
changes to a nonmonotonous plot (with one negative global minimum), 
observed at $\eta=1$.
The answer to this question is presented for the coupling $\Gamma=6$
and $N=6$ particles in Fig. \ref{Fig:5}.
It turns out that as soon as $\eta$ is nonzero $\tilde{P}$ exhibits 
a nonmonotonous behavior with one negative (global) minimum. 
Consequently, the necessary and sufficient condition for the attraction 
phenomenon is the presence of a nonzero charge density on both walls.
We suggest that the same condition applies to the analogous 3D models 
with counterions only. 

\begin{figure}[tbh]
\begin{center}
\includegraphics[width=0.8\textwidth,clip]{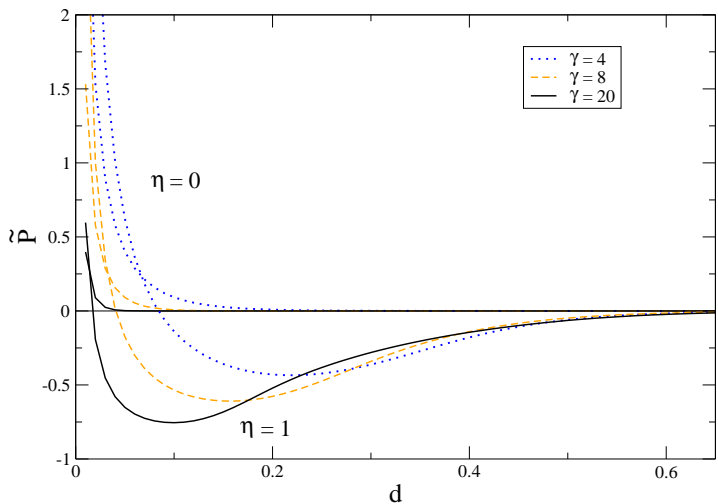}
\caption{The pressure $\tilde{P}$ versus the distance between the walls $d$ 
for $N=2$ particles. 
The dotted (blue) curves correspond to $\gamma=4$, the dashed (orange) curves 
to $\gamma=8$ and the solid (black) curves to $\gamma=20$.
The curves located only above the $d$-axis are evaluated with the asymmetry 
parameter $\eta=0$ and those going also below the $d$-axis with $\eta=1$.}
\label{Fig:6}
\end{center}
\end{figure}

The formula for $N=2$ particle system (\ref{Q2}), valid for any integer
$\gamma$, is used in Fig. \ref{Fig:6} to visualize the effect of increasing 
the coupling on the dependence of the pressure $\tilde{P}$ on the distance $d$.
The chosen values of $\gamma$ are $4$ (dotted curve), $8$ (dashed curve) and 
$20$ (solid curve).
For the asymmetry parameter $\eta=0$, all plots decay monotonously to 0
at $d\to\infty$, as is expected from the previous analysis.
For $\eta=1$, all plots exhibit a global (negative) minimum and goes to
0 at $d\to\infty$ from below.  
It is seen that by increasing $\gamma$ the global minimum of $\tilde{P}$ 
goes down; for $\gamma=20$ it approaches the lower bound $-1$. 

\renewcommand{\theequation}{7.\arabic{equation}}
\setcounter{equation}{0}

\section{Conclusion} \label{Sec:7}
The neutral system of identical pointlike charges moving on the surface of 
a cylinder of circumference $W$ and length $d$, with the like-charged
(symmetrically or asymmetrically) end-circles, was of interest in 
this paper.  
Like any 2D one-component model, it admits a 1D anticommuting-field
representation which permits one to express the one-body, two-body, etc.
densities of particles in terms of the anticommuting-field correlators.
Specific transformations of the anticommuting variables, which preserve
the composite form of the operators (\ref{composite}), imply specific
sum rules for the statistical quantities which are basically of two types.
The sum rules (\ref{zerothmoment}) and (\ref{Wertheim}) are the obvious
finite-$W$ generalizations of the 2D zeroth-moment Stillinger-Lovett
and WLMB conditions, respectively.
Another sum rules, the one given by Eqs. (\ref{result1}), (\ref{dodatok1})
and the other given by Eqs. (\ref{result1}), (\ref{dodatok1}), provide
in the limit $W\to\infty$ the new exact constraints (\ref{sumrule2})
and (\ref{sepa}) for the prefactor function $f^{(2)}(x,x')$ of the 
asymptotic behavior of the Ursell function along the two walls 
(\ref{asymptotic2}).

The possibility of an effective attraction between like-charged walls
was another important subject investigated in this paper. 
The exactly solvable case of the free-fermion coupling $\Gamma=2$ was 
studied in Sect. \ref{Sec:5}.
For the symmetrically charged walls $(\eta=1)$, the monotonous dependence of 
the (dimensionless) pressure $\tilde{P}$ on the (dimensionless) distance
between the walls $d$ in Fig. \ref{Fig:2} shows that the results for $N=8$ and 
$N\to\infty$ particles are practically indistinguishable.
We expect that also for higher values of $\Gamma$ the results for $N=8$
particles describe adequately those in the thermodynamic limit.
For $\Gamma=4$ and $6$ we were able to derive the exact expressions for 
the dependence $\tilde{P}(d)$ up to $N=8$ particles, 
see Figs. \ref{Fig:3} and \ref{Fig:4}.
If there is no charge on one of the walls $(\eta=0)$, the pressure $\tilde{P}$ 
is always positive and decreases monotonously with the distance $d$.
In the case of the symmetrically charged walls $(\eta=1)$, the plot
of $\tilde{P}$ versus $d$ has one global minimum and goes to 0 at $d\to\infty$
from below.
In other words, the repulsion between the walls in the region of small $d$
changes at a specific distance to the attraction which lasts up to $d\to\infty$.
Note a small difference between data for $N=2$ and $N=8$. 
As is shown in Fig. \ref{Fig:5} for $\Gamma=6$ and $N=6$, the change from
the monotonous to nonmonotonous behavior of $\tilde{P}$ occurs as soon as 
$\eta>0$, i.e., when the two walls are like-charged by a nonzero line charge 
density the attraction takes place at sufficiently large distances.
The fact that the attraction phenomena between like-charged lines
occurs starting from a relatively small coupling, somewhere between $\Gamma=2$
and $\Gamma=4$, is surprising.

As concerns our future plans, it might be interesting to extend the present
analysis to other one-component Coulomb systems like the jellium model.

\begin{acknowledgement}
The support received from the project EXSES APVV-16-0186 and VEGA Grant
No. 2/0003/18 is acknowledged.
\end{acknowledgement}


\begin{thebibliography}{10}

\bibitem{Alastuey81} Alastuey, A., Jancovici, B.: 
On the classical two-dimensional one-component Coulomb plasma.
J. Physique {\bf 42}, 1--12 (1981)

\bibitem{Andelman06} Andelman, D.:
Introduction to electrostatics in soft and biological matter. 
In: Poon, W.C.K., Andelman, D. (eds.) Soft Condensed Matter Physics
in Molecular and Cell Biology, vol. 6. Taylor \& Francis, New York (2006)

\bibitem{Attard88} Attard, P., Mitchell, D.J., Ninham, B.W.:
Beyond Poisson-Boltzmann: Images and correlations in the electric double
layer. I. Counterions only.
J. Chem. Phys. {\bf 88}, 4987--4996  (1988)

\bibitem{Attard96} Attard, Ph.:
Electrolytes and the electric double layer.
Adv. Chem. Phys. {\bf XCII}, 1--159 (1996)

\bibitem{Bakhshandeh11} Bakhshandeh, A., dos Santos, A.P., Levin, Y.: 
Weak and strong coupling theories for polarizable colloids and nanoparticles.
Phys. Rev. Lett. {\bf 107}, 107801 (2011)

\bibitem{Barbosa00} Barbosa, M.C., Deserno, M., Holm, C.: 
A stable local density functional approach to ion-ion correlations.
Europhys. Lett. {\bf 52}, 80--86 (2000)

\bibitem{Baus80} Baus, M., Hansen, J.P.: 
Statistical mechanics of simple Coulomb systems.
Phys. Rep. {\bf 59}, 1--94 (1980)

\bibitem{Baxter63} Baxter, R.J.:
Statistical mechanics of a one-dimensional Coulomb system with 
a uniform charge background.
Proc. Camb. Phil. Soc. {\bf 59}, 779--787 (1963)

\bibitem{Ben11} Ben-Yaakov, D., Andelman, D., Podgornik, R., Harries, D.:
Ion-specific hydration effects: Extending the Poisson-Boltzmann theory.
Curr. Opin. Colloid. Interface Sci. {\bf 16}, 542--550 (2011)

\bibitem{Bloomfield91} Bloomfield, V.A.: 
Condensation of DNA by multivalent cations: Considerations on mechanism.
Biopolymers {\bf 31}, 1471--1481 (1991)

\bibitem{Blum81} Blum, L., Henderson, D., Lebowitz, J.L., Gruber, Ch., 
Martin, Ph.A.: 
A sum rule for an inhomogeneous electrolyte.
J. Chem. Phys. {\bf 75}, 5974--5975 (1981) 

\bibitem{Boroudjerdi05} Boroudjerdi, H., Kim, Y.-W., Naji, A., Netz, R.R., 
Schlagberger, X., Serr, A.:
Statics and dynamics of strongly charged soft matter.
Phys. Rep. {\bf 416}, 129-199 (2005)

\bibitem{Bratko86} Bratko, D., J\"onsson, B., Wennerstr\"om, H.: 
Electrical double layer interactions with image charges.
Chem. Phys. Lett. {\bf 128}, 449--454 (1986)

\bibitem{Burak04} Burak, Y., Andelman, D.: 
Test-charge theory for the electric double layer.
Phys. Rev. E {\bf 70}, 016102 (2004)

\bibitem{Burak06} Burak, Y., Orland, H.:
Manning condensation in two dimensions.
Phys. Rev. E {\bf 73}, 010501(R) (2006)

\bibitem{Carnie81} Carnie, S.L., Chan, D.Y.C.: 
The Stillinger-Lovett condition for non-uniform electrolytes.
Chem. Phys. Lett. {\bf 77}, 437--440 (1981)

\bibitem{Choquard80} Choquard, Ph., Favre, P., Gruber, Ch.: 
On the equation of state of classical one component systems with long range
forces.
J. Stat. Phys. {\bf 23}, 405--442 (1980)

\bibitem{Choquard81} Choquard, Ph.: 
The two-dimensional one component plasma on a periodic strip.
Helv. Phys. Acta {\bf 54}, 332--332 (1981)  

\bibitem{Dean98} Dean, D.S., Horgan, R.R., Sentenac, D.:
Boundary effects in the one-dimensional Coulomb gas.
J. Stat. Phys. {\bf 90}, 899–-926 (1998)

\bibitem{Dean09} Dean, D.S., Horgan, R.R., Naji, A., Podgornik, R.: 
One-dimensional counterion gas between charged surfaces: 
Exact results compared with weak- and strong-coupling analyses.
J. Chem. Phys. {\bf 130}, 094504 (2009)

\bibitem{Dubois98} Dubois, M., Zemb, T., Fuller, N., Rand, R.P., 
Pargesian, V.A.: 
Equation of state of a charged bilayer system: Measure of the entropy of 
the lamellar–lamellar transition in DDABr. 
J. Chem. Phys. {\bf 108}, 7855--7869 (1998)

\bibitem{Edwards62} Edwards, S.F., Lenard, A.: Exact statistical mechanics of 
a one‐dimensional system with Coulomb forces. II. The method of functional 
integration. J. Math. Phys. {\bf 3}, 778--792 (1962)

\bibitem{Forrester98} Forrester, P.J.: 
Exact results for two-dimensional Coulomb systems.
Phys. Rep. {\bf 301}, 235--270 (1998)   

\bibitem{Forsman04} Forsman, J.: 
A simple correlation-corrected Poisson-Boltzmann theory.
J. Phys. Chem. B {\bf 108}, 9236--9245 (2004)

\bibitem{Goldoni96} Goldoni, G., Peeters, M.:
Stability, dynamical properties, and melting of a classical bilayer Wigner 
crystal.
Phys. Rev. B {\bf 53}, 4591--4603 (1996)

\bibitem{Grimaldo15} Grimaldo, J.A.M., T\'ellez, G.: 
Relations among two methods for computing the partition function of the 
two-dimensional one-component plasma.
J. Stat. Phys. {\bf 160}, 4--28 (2015)

\bibitem{Gronbech97} Gr{\o}nbech-Jensen, N., Mashl, R.J., Bruinsma, R.F.,
Gelbart, W.M.: 
Counterion-Induced attraction between rigid polyelectrolytes.
Phys. Rev. Lett. {\bf 78}, 2477--2480 (1997) 

\bibitem{Grosberg02} Grosberg, A.Y., Nguyen, T.T., Shklovskii, B.I.: 
Colloquium: The physics of charge inversion in chemical and biological systems.
Rev. Mod. Phys. {\bf 74}, 329--345 (2002)

\bibitem{Gulbrand84} Gulbrand, L., J\"onsson, B., Wennerstr\"om, H., Linse, P.: 
Electrical double layer forces. A Monte Carlo study.
J. Chem. Phys. {\bf 80}, 2221-2228 (1984)

\bibitem{Hansen00} Hansen, J.P., L\"owen, H.:
Effective interactions between electric double layers.
Annu. Rev. Phys. Chem. {\bf 51}, 209--242 (2000) 

\bibitem{Henderson78} Henderson, D., Blum, L.: 
Some exact results and the application of the mean spherical approximation 
to charged hard spheres near a charged hard wall.
J. Chem. Phys. {\bf 69}, 5441--5449 (1978)

\bibitem{Henderson79} Henderson, D., Blum, L., Lebowitz, J.L.: 
An exact formula for the contact value of the density profile of 
a system of charged hard spheres near a charged wall.
J. Electroanal. Chem. {\bf 102}, 315--319 (1979)

\bibitem{Jancovici81} Jancovici, B.: 
Exact results for the two-dimensional one-component plasma.
Phys. Rev. Lett. {\bf 46}, 386--388 (1981)

\bibitem{Jancovici82a} Jancovici, B.:
Classical Coulomb systems near a plane wall. I.
J. Stat. Phys. {\bf 28}, 43--65 (1982)

\bibitem{Jancovici82b} Jancovici, B.:
Classical Coulomb systems near a plane wall. II.
J. Stat. Phys. {\bf 29}, 263--280 (1982)

\bibitem{Jancovici84} Jancovici, B.:
Surface properties of a classical two-dimensional one-component plasma: 
Exact results.
J. Stat. Phys. {\bf 34}, 803--815 (1984)

\bibitem{Jancovici92} Jancovici, B.:
Inhomogeneous two-dimensional plasmas.
In: Henderson. D. (ed.) Inhomogeneous Fluids, pp. 201-237, Dekker, 
New York (1992)

\bibitem{Jancovici95} Jancovici, B.:
Classical Coulomb systems: Screening and correlations revisited.
J. Stat. Phys. {\bf 80}, 445--459 (1995)

\bibitem{Jancovici01} Jancovici, B., \v{S}amaj, L.:
Charge correlations in a Coulomb system along a plane wall:
A relation between asymptotic behavior and dipole moment.
J. Stat. Phys. {\bf 105}, 193--209 (2001)

\bibitem{Kanduc07} Kandu\v{c}, M., Podgornik, R.: 
Electrostatic image effects for counterions between charged planar walls.
Eur. Phys. J. E {\bf 23}, 265--274 (2007)

\bibitem{Kanduc08} Kandu\v{c}, M., Trulsson, M., Naji, A., Burak, Y., 
Forsman, J., Podgornik, R.: 
Weak- and strong-coupling electrostatic interactions between asymmetrically
charged planar surfaces.
Phys. Rev. E {\bf 78}, 061105 (2008)

\bibitem{Kanduc10} Kandu\v{c}, M., Naji, Forsman, J., Podgornik, R.: 
Dressed counterions: Strong electrostatic coupling in the presence of salt.
J. Chem. Phys. {\bf 132}, 124701 (2010)

\bibitem{Kanduc11} Kandu\v{c}, M., Naji, Forsman, J., Podgornik, R.: 
Dressed counterions: Polyvalent and monovalent ions at charged dielectric
interfaces.
Phys. Rev. E {\bf 84}, 011502 (2011)

\bibitem{Kanduc12} Kandu\v{c}, M., Naji, Forsman, J., Podgornik, R.: 
Attraction between neutral dielectrics mediated by multivalent ions in 
an asymmetric ionic fluid.
J. Chem. Phys. {\bf 137}, 174704 (2012)

\bibitem{Kanduc17} Kandu\v{c}, M., Moazzami-Gudarzi, M., Valmacco, V.,
Podgornik, R., Trefalt, G.: Interactions between charged particles with 
bathing multivalent counterions: experiments vs. dressed ion theory .
Phys. Chem. Chem. Phys. {\bf 19}, 10069--10080 (2017)

\bibitem{Kekicheff93} K\'ekicheff, P., Mar\v{c}elja, S., Senden, T.J., 
Shubin, V.E.: 
Charge reversal seen in electrical double layer interaction of surfaces 
immersed in 2:1 calcium electrolyte. 
J. Chem. Phys. {\bf 99}, 6098--6113 (1993)

\bibitem{Khan85} Khan, A., J\"onsson, B., Wennerstr\"om, H.: 
Phase equilibria in the mixed sodium and calcium di-2-ethylhexylsulfosuccinate 
aqueous system. An illustration of repulsive and attractive double-layer forces.
J. Phys. Chem. {\bf 89}, 5180-5184 1985

\bibitem{Kjellander84} Kjellander, R., Mar\v{c}elja, S.: 
Correlation and image charge effects in electric double-layers.
Chem. Phys. Lett. {\bf 112}, 49--53 (1984)

\bibitem{Kjellander88} Kjellander, R., Mar\v{c}elja, S., Quirk, J.P.: 
Attractive double-layer interactions between calcium clay particles.
J. Colloid Interface Sci. {\bf 126}, 194--211 (1988)

\bibitem{Kunz74} Kunz, H.:
The one-dimensional classical electron gas.
Ann. Phys. {\bf 85}, 303--335 (1974)

\bibitem{Lenard61} Lenard, A.:
Exact statistical mechanics of a one‐dimensional system with Coulomb forces. 
J. Math. Phys. {\bf 2}, 682--693 (1961)

\bibitem{Levin99} Levin, Y., Arenzon, J.J., Stilck, J.F.: 
The nature of attraction between like-charged rods.
Phys. Rev. Lett. {\bf 83}, 2680 (1999)

\bibitem{Levin02} Levin, Y.:
Electrostatic correlations: from Plasma to Biology.
Rep. Prog. Phys. {\bf 65}, 1577 (2002)

\bibitem{Lovett76} Lovett, R., Mou, C.Y., Buff, F.P.: 
The structure of the liquid-vapor interface.
J. Chem. Phys. {\bf 65}, 570--572 (1976)

\bibitem{Ma} Ma, Sh.-K.:
Statistical mechanics. World Scientific, Singapore (1985)

\bibitem{Martin88} Martin, Ph.A.:
Sum rules in charged fluids.
Rev. Mod. Phys. {\bf 60}, 1075--1127 (1988)

\bibitem{Mehta} Mehta, M.L.:
Random Matrices. 2nd ed. Academic Press, London (1990)

\bibitem{Messina09} Messina, R.:
Electrostatics in soft matter.
J. Phys.: Condens. Matter {\bf 21}, 113102 (2009)

\bibitem{Moreira00} Moreira, A.G., Netz, R.R.: 
Strong-coupling theory for counter-ion distributions.
Europhys. Lett. {\bf 52}, 705--711 (2000)

\bibitem{Moreira01} Moreira, A.G., Netz, R.R.: 
Binding of similarly charged plates with counterions only.
Phys. Rev. Lett. {\bf 87}, 078301 (2001)

\bibitem{Moreira02} Moreira, A.G., Netz, R.R.: 
Simulations of counterions at charged plates.
Eur. Phys. J. E {\bf 8}, 33--58 (2002)

\bibitem{Naji05} Naji, A., Netz, R.R.: 
Counterions at charged cylinders: Criticality and universality 
beyond mean-field theory.
Phys. Rev. Lett {\bf 95}, 185703 (2005)

\bibitem{Naji06} Naji, A., Netz, R.R.:
Scaling and universality in the counterion-condensation transition at 
charged cylinders.
Phys. Rev. E {\bf 73}, 056105 (2006)

\bibitem{Naji13} Naji, A., Kandu\v{c}, M., Forsman, J., Podgornik, R.:
Perspective: Coulomb fluids -- Weak coupling, strong coupling, in between and 
beyond. 
J. Chem. Phys. {\bf 139}, 150901 (2013)

\bibitem{Netz00} Netz, R.R., Orland, H.:
Beyond Poisson-Boltzmann: Fluctuation effects and correlation functions.
Eur. Phys. J. E {\bf 1}, 203--214 (2000)

\bibitem{Netz01} Netz, R.R.:
Electrostatics of counter-ions at and between planar charged walls: from
Poisson-Boltzmann to the strong-coupling theory.
Eur. Phys. J. E {\bf 5}, 557--574 (2001)

\bibitem{Nordholm84} Nordholm, S.: 
Simple analysis of the thermodynamic properties of the one-component plasma.
Chem. Phys. Lett. {\bf 105}, 302--307 (1984)

\bibitem{Palaia18} Palaia, I., Trulsson, M., \v{S}amaj, L., Trizac, E.: 
A correlation-hole approach to the electric double layer with counter-ions
only. Mol. Phys. {\bf 116}, 3134--3146 (2018)

\bibitem{Podgornik90} Podgornik, R.:
An analytic treatment of the first-order correction to the Poisson-Boltzmann
interaction free energy in the case of counter-ion only Coulomb fluid.
J. Phys. A: Math. Gen. {\bf 23}, 275--284 (1990)

\bibitem{Rau92} Rau, D.C., Pargesian, V.A.: 
Direct measurement of the intermolecular forces between counterion-condensed 
DNA double helices. Evidence for long range attractive hydration forces.
Biophys. J. {\bf 61}, 246--259 (1992)

\bibitem{Rouzina96} Rouzina, I., Bloomfield, V.A.: 
Macroion attraction due to electrostatic correlation between screening 
counterions. 1. Mobile surface-adsorbed ions and diffuse ion cloud.
J. Phys. Chem. {\bf 100}, 9977--9989 (1996)

\bibitem{Samaj95} \v{S}amaj, L., Percus, J.K.:
A functional relation among the pair correlations of the two-dimensional
one-component plasma.
J. Stat. Phys. {\bf 80}, 811--824 (1995)

\bibitem{Samaj00} \v{S}amaj, L.:
Microscopic calculation of the dielectric susceptibility tensor for 
Coulomb fluids.
J. Stat. Phys. {\bf 100}, 949--967 (2000)

\bibitem{Samaj04a} \v{S}amaj, L.:
Is the two-dimensional one-component plasma exactly solvable?
J. Stat. Phys. {\bf 117}, 131--158 (2004)

\bibitem{Samaj04b} \v{S}amaj, L., Wagner, J., Kalinay, P.: 
Translation symmetry breaking in the one-component plasma on the cylinder.
J. Stat. Phys. {\bf 117}, 159--178 (2004)

\bibitem{Samaj11a} \v{S}amaj, L., Trizac, E.:
Counterions at highly charged interfaces: From one plate to like-charge
attraction.
Phys. Rev. Lett. {\bf 106}, 078301 (2011)

\bibitem{Samaj11b} \v{S}amaj, L., Trizac, E.: 
Wigner-crystal formulation of strong-coupling theory for counterions
near planar charged interfaces.
Phys. Rev. E {\bf 24}, 041401 (2011).

\bibitem{Samaj11c} \v{S}amaj, L., Trizac, E.:
Counter-ions at charged walls: Two-dimensional systems.
Eur. Phys. J. E {\bf 34}, 20 (2011)

\bibitem{Samaj13} \v{S}amaj, L.:
Counter-ions at single charged wall: Sum rules.
Eur. Phys. J. E {\bf 36}, 100 (2013)

\bibitem{Samaj14} \v{S}amaj, L., Trizac, E.:
Counter-ions between or at asymmetrically charged walls: 2D free-fermion point.
J. Stat. Phys. {\bf 156}, 932--947 (2014)

\bibitem{Samaj15} \v{S}amaj, L.:
Counter-ions near a charged wall: Exact results for disc and planar geometries.
J. Stat. Phys. {\bf 161}, 227--249 (2015)

\bibitem{Samaj16a} \v{S}amaj, L., dos Santos, A.P., Levin, Y., Trizac, E.: 
Mean-field beyond mean-field: the single particle view for moderately 
to strongly coupled charged fluids.
Soft Matter {\bf 12}, 8768--8773 (2016)

\bibitem{Samaj16b} \v{S}amaj, L.: 
Amplitude function of asymptotic correlations along charged wall in 
Coulomb fluids.
J. Stat. Phys. {\bf 164}, 304--320 (2016)

\bibitem{Samaj17} \v{S}amaj, L.: 
Finite-size effects in non-neutral two-dimensional Coulomb fluids.
J. Stat. Phys. {\bf 168}, 434--446 (2017)

\bibitem{Samin11} Samin, S., Tsori, I.:
Attraction between like-charge surfaces in polar mixtures.
Europhys. Lett. {\bf 95}, 36002 (2011) 

\bibitem{Santangelo06} Santangelo, C.D.:
Computing counterion densities at intermediate coupling.
Phys. Rev. E {\bf 73}, 041512 (2006)

\bibitem{Schweigert99} Schweigert, I.V., Schweigert, V.A., Peeters, F.M.:
Melting of the classical bilayer Wigner crystal: Influence of lattice symmetry.
Phys. Rev. Lett. {\bf 82}, 5293--5296 (1999)

\bibitem{Shklovskii99} Shklovskii, B.I.: 
Screening of a macroion by multivalent ions: 
Correlation-induced inversion of charge.
Phys. Rev. E {\bf 60}, 5802--5811 (1999)

\bibitem{Stillinger68a} Stillinger, F.H., Lovett, R.: 
Ion-pair theory of concentrated electrolytes. I. Basic concepts.
J. Chem. Phys. {\bf 48}, 3858--3868 (1968)

\bibitem{Stillinger68b} Stillinger, F.H., Lovett, R.: 
General restriction on the distribution of ions in electrolytes.
J. Chem. Phys. {\bf 49}, 1991--1994 (1968)

\bibitem{Strandburg88} Strandburg, K.J.: 
Two-dimensional melting.
Rev. Mod. Phys. {\bf 60}, 161--207 (1988)

\bibitem{Tellez99} T\'ellez, G., Forrester, P.J.: 
Exact finite-size study of the 2d-OCP at $\Gamma=4$ and $\Gamma=6$. 
J. Stat. Phys. {\bf 97}, 489--521 (1999)

\bibitem{Tellez12} T\'ellez, G., Forrester, P.J.: 
Expanded Vandermonde powers and sum rules for the two-dimensional 
one-component plasma. 
J. Stat. Phys. {\bf 147}, 825--855 (2012)

\bibitem{Tellez15} T\'ellez, G., Trizac, E.: 
Screening like charges in one-dimensional Coulomb systems: Exact results. 
Phys. Rev. E {\bf 92}, 042134 (2015)

\bibitem{Usenko79} Usenko, A.S., Yakimenko, I.P.:
Interaction energy of stationary charges in a bounded plasma.
Sov. Tech. Phys. Lett. {\bf 5}, 549--550 (1979) 

\bibitem{Varela17} Varela, L., T\'ellez, G., Trizac, E.: 
Configurational and energy landscape in one-dimensional Coulomb systems.
Phys. Rev. E {\bf 97}, 022112 (2017)

\bibitem{Wertheim76} Wertheim, M.S.: 
Correlations in the liquid-vapor interface.
J. Chem. Phys. {\bf 65}, 2377--2381 (1976)

\end{thebibliography}
\end{document}